\documentclass[prl,twocolumn]{revtex4}
\usepackage{amsfonts}
\usepackage{mathrsfs}
\usepackage{amsmath}
\usepackage{amssymb}
\usepackage[medium]{titlesec}
\usepackage{bm}
\usepackage[normalem]{ulem}
\usepackage{extarrows}
\usepackage{slashed}
\usepackage{isodateo}
\usepackage{graphicx} 
\usepackage{xcolor}
\usepackage[utf8]{inputenc}
\usepackage[bookmarksnumbered=true,bookmarksopen=true]{hyperref}
 \hypersetup{colorlinks,%
             linkcolor=[rgb]{0,0.3,0.6}, %
             citecolor=[rgb]{0,0.3,0.6}, %
             urlcolor=[rgb]{0,0.3,0.6}}
\usepackage{indentfirst}
\renewcommand{\FR}[2]{\displaystyle\frac{\,{#1}\,}{#2}}
\newcommand{\fr}[2]{\mbox{$\frac{\,{#1}\,}{#2}$}}

\graphicspath{{fig/}}

\usepackage[eulergreek]{sansmath}

\def\bge{\begin{equation}}
\def\ede{\end{equation}}
\def\bga{\begin{aligned}}
\def\eda{\end{aligned}}
\def\bgp{\begin{pmatrix}}
\def\edp{\end{pmatrix}}
\def\bgs{\begin{subequations}}
\def\eds{\end{subequations}}
\newcommand{\order}[1]{\mathcal{O}({#1})}
\def\di{{\mathrm{d}}}

\def\mb{\mathbf}

\def\ra{\rangle}

\def\al{\alpha}

\def\ga{\gamma}
\def\de{\delta}

\def\ka{\kappa}

\def\si{\sigma}

\newcommand{\ob}[1]{\mkern 2mu \overline{\mkern -2mu #1 \mkern -2mu}\mkern 2mu}

\newcommand{\wh}[1]{\mkern 2mu \widehat{\mkern-2mu#1\mkern-2mu}\mkern 2mu}

\begin{document} 

\title{Probing Ultralight Bosons with Compact Eccentric Binaries}

\author{Boye Su}
\email{sby20@mails.tsinghua.edu.cn}
\affiliation{Department of Physics, Tsinghua University, Beijing 100084, China} 

\author{Zhong-Zhi Xianyu}
\email{zxianyu@tsinghua.edu.cn}
\affiliation{Department of Physics, Tsinghua University, Beijing 100084, China}

\author{Xingyu Zhang}
\email{zhang-xy19@mails.tsinghua.edu.cn}
\affiliation{Department of Physics, Tsinghua University, Beijing 100084, China}  
 
\begin{abstract}

Ultralight bosons can be abundantly produced through superradiance process by a spinning black hole and form a bound state with hydrogen-like spectrum. We show that such a ``gravitational atom'' typically possesses anomalously large mass quadrupole and leads to significant orbital precession when it forms an eccentric binary with a second compact object. Dynamically formed black hole binaries or pulsar-black hole binaries are typically eccentric during their early inspirals. We show that the large orbital precession can generate distinct and observable signature in their gravitational wave or pulsar timing signals.
\end{abstract}

\maketitle

A new era of gravitational wave (GW) astronomy has emerged following the LIGO discovery of two coalescing black holes (BHs) \cite{LIGOScientific:2016aoc}. Since then it has been extensively discussed how we can use this new opportunity to probe fundamental physics beyond the Standard Model (SM) \cite{Perkins:2020tra}. Among many ideas put forward, using the compact binaries to search for new light degrees of freedom has attracted lots of attentions.
 
Light bosons are ubiquitous in new physics models, and can be good dark matter candidates \cite{Essig:2013lka}. They are typically very weakly coupled to SM and have escaped all direct experimental searches so far. However, they most likely gravitate as known particles do. Strong gravity systems like BHs are thus natural laboratories to study them. In particular, it is known that a fast rotating BH of mass $M$ can copiously produce light bosons of mass $\nu$ due to superradiance instability, a pure gravitational effect \cite{Zeldovich1971,Starobinsky:1973aij,Brito:2015oca}. When the ``gravitational fine structure constant'' $\al\equiv GM\nu/\hbar c \ll 1$, the boson $\si$ can form nonrelativistic and metastable condensate around the BH with hydrogen-like spectrum. We can thus call the formed condensate a ``superradiance cloud'' (SC) and call the whole system of SC + central BH a ``gravitational atom'' (GA). For stellar mass BHs, the corresponding mass of $\si$ is around $\nu\sim 10^{-13}\text{eV}(\al/0.01)(10M_\odot/M)$.

The phenomenology of SC has long been under active study. See e.g.\ \cite{Arvanitaki:2010sy,Cardoso:2011xi,Yoshino:2012kn,Arvanitaki:2014wva}. The LIGO/Virgo discovery of many coalescing BH binaries added new opportunities \cite{LIGOScientific:2018mvr,LIGOScientific:2020ibl}. Recent studies have exploited the inspiral phase of a GA-borne binary, showing that a resonant level transition can occur if the binary’s orbital period matches the energy split of certain levels of a GA \cite{Baumann:2018vus,Baumann:2019eav,Baumann:2019ztm,Ding:2020bnl,Tong:2021whq}. This is dubbed “gravitational collider physics” \cite{Baumann:2019ztm} as the induced resonant transition is reminiscent of a resonant peak of on-shell particle production at a real collider.

The level transition requires the match of frequencies of two different origins, which is expected to be rare in generic inspiral systems. Furthermore, the GW signals of level transition, while being a distinct signature, could be challenging to search for in realistic GW experiments. Therefore it will be useful to look for alternative signals of GAs that are less restrictive on binary’s orbital parameters and easier to search for.

In this Letter, we point out such a signal for binaries with eccentric orbits. The key point is that the GA typically possesses huge mass quadrupole Q. In terms of a dimensionless quadrupole parameter $\ka\equiv-QM/J^2$ (where $J$ is the spin), a GA can have $\ka\sim10^3(0.1/\al)^3$. For comparison, $\ka=1$ for Kerr BHs and $\ka\sim\order{1\text{-}10}$ for a neutron star \cite{Laarakkers:1997hb,Pappas:2012ns}. As a result, a GA can induce very significant apsidal precession of an eccentric orbit for a wide range of orbital parameters, which is the main signal of this Letter. We will show that this signal can be very significant and potentially observable for BH-GA binaries and pulsar-GA binaries.

\begin{figure}
\centering
\includegraphics[width=0.38\textwidth]{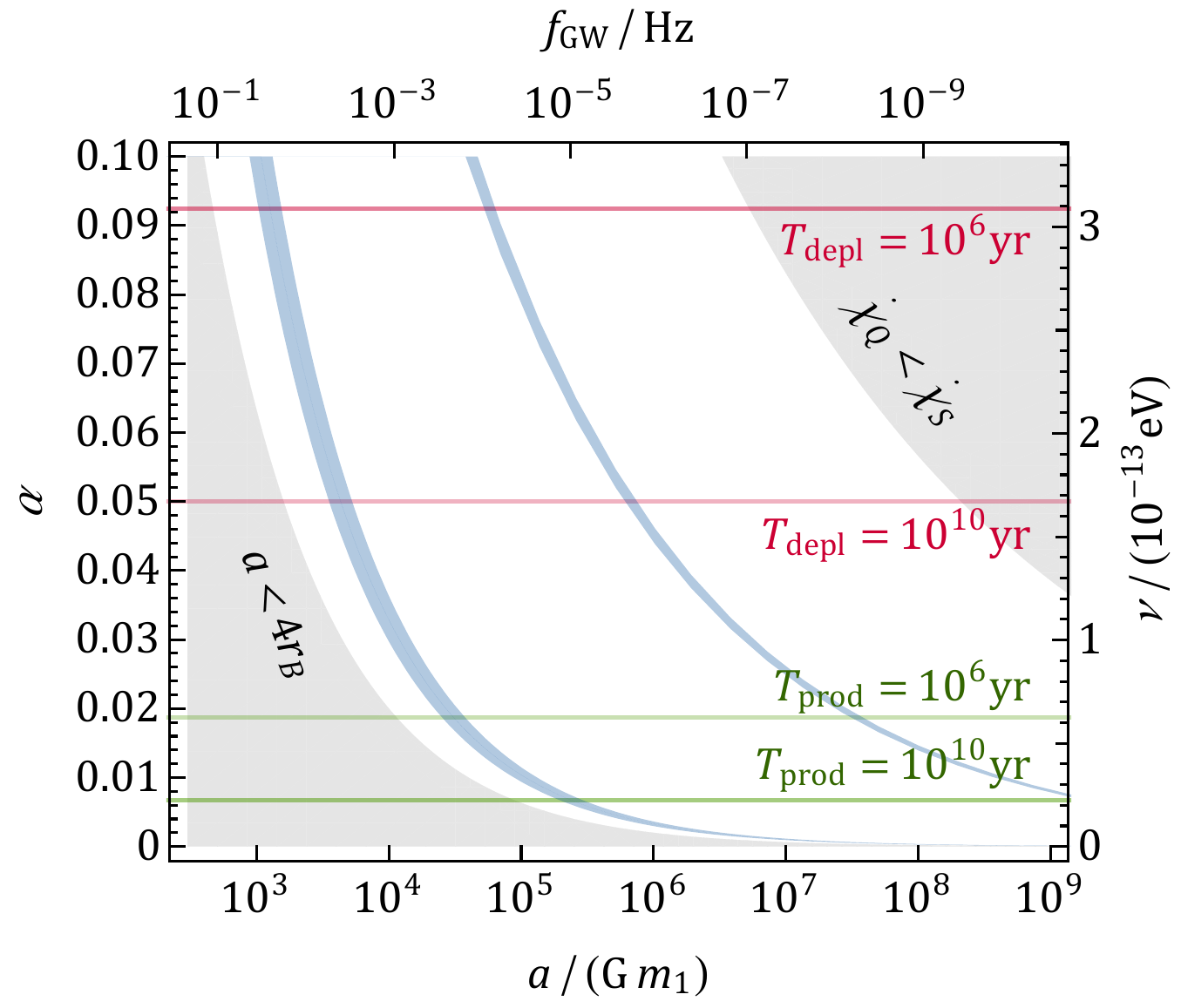}
\caption{The allowed region (unshaded) for observing $\dot\chi_\text{Q}$ in $(\al,a)$ plane, for a binary of $(40+40)M_\odot$, assuming saturated $|211\ra$ state. The two shaded regions in gray are excluded by the condition $a>4r_\text{B}$ and $\dot\chi_\text{Q}>\dot\chi_\text{S}$, respectively. The blue belts mark several resonant transition bands \cite{Baumann:2019ztm}.}
\label{fig_alphaDist}
\end{figure}

Our signal does not rely on any resonant transitions, and therefore can be viewed as an “off-shell” signal of gravitational collider physics. Thus it shares both the merit and the drawback of “off-shell” observables. It shows up for wider range of parameters, but it does not probe the internal structure of a GA. The signal can nevertheless be a useful probe of GAs, since we are virtually unaware of any other astrophysical objects sharing similar masses and mass quadrupoles. Thus the apsidal precession signal opens up a continuous and large parameter space for detecting GA beyond the known narrow resonance band, shown in Fig.\;\ref{fig_alphaDist}.

LIGO-type binaries are expected to have observably large eccentricity at low frequencies, as predicted by a wide class of dynamical formation scenarios \cite{Samsing:2013kua,Samsing:2017xmd,Samsing:2018isx,Samsing:2019dtb,Antonini:2012ad,Hoang:2017fvh,Fragione:2019dtr,Randall:2017jop,Randall:2018nud,Randall:2018lnh,Silsbee:2016djf,Randall:2019sab,Deme:2020ewx}. Such eccentric binaries are important targets of future space GW telescopes \cite{Ruan:2018tsw,Liu:2020eko,Toubiana:2020cqv,Buscicchio:2021dph,Randall:2021xjy}. On the other hand, known pulsar binaries typically possess nonzero eccentricity. No pulsar-BH binaries are detected yet but they could be found in the future \cite{Faucher-Giguere:2010dol,Shao:2018qpt}. Therefore we shall consider our signal in eccentric pulsar-BH binaries as well.

The effects of the mass quadrupole on binary’s orbit and corresponding GWs have long been studied \cite{Barker:1975ae,Poisson:1997ha,Barack:2006pq}. But to our best knowledge, our work is the first to address GA-induced apsidal precession in eccentric BH-BH and pulsar-BH binaries. See \cite{Krishnendu:2017shb,Ferreira:2017pth,Hannuksela:2018izj,Zhang:2019eid,GRAVITY:2019tuf} for related works.

\noindent{\bf Superradiance cloud.} In this Letter we consider a real scalar field $\si$ of mass $\nu$. Generalization to other bosonic species will be presented in a companion paper \cite{preclong}. The superradiance and the GA formation can be studied by solving the field equation $(\square-\nu^2)\si=0$ outside a Kerr
BH of mass $M$ and spin $J$. Here and below we take $\hbar=c=1$. When $\al\ll1$, one finds hydrogen-like bound states $|\mathsf{n\ell m}\ra$. At leading order in $\al$ the binding energy of $|\mathsf{n\ell m}\ra$ is given by $E_\mathsf{n} =-\al^2 \nu /(2\mathsf{n}^2 )\ll \nu$ , justifying the nonrelativistic approximation.

One important difference from the hydrogen, though, is the ingoing boundary condition for $\si$ at the BH horizon, which introduces imaginary parts in the energy eigenvalues and thus can trigger production or depletion of states. Detailed study shows that $|211\ra$ and $|322\ra$ are the two leading states that a spinning BH can produce which then remain quasi-stable. The leading depletion channel for them is GW radiation, which is typically very slow \cite{Yoshino:2013ofa}. Therefore, these states can exist with astronomical lifetime. Other states might also be relevant but we will mainly focus on $|211\ra$ below.
  
The production and depletion time scales for $|211\ra$ have been worked out as (See e.g.\ \cite{Baumann:2019ztm})
\begin{align}
&T_\text{prod}(|211\ra)\sim10^6\text{yr}(M/10M_\odot)(0.016/\al)^9,\\
&T_\text{depl}(|211\ra)\sim10^8\text{yr}(M/10M_\odot)(0.062/\al)^{15}.
\end{align}

This puts bounds on $\al$ for given masses, shown in Fig.\;\ref{fig_alphaDist}. After a period of time $T_\text{prod}$, a SC is formed in $|211\ra$ state. Its energy density at leading order in $\al$ is time independent,
\bge
  \rho_{211}(\mb x)=\frac{1}{64\pi}\al^5\nu^5m_\text{C}\,r^2 e^{-\al\nu r}\sin^2\theta,
\ede
where $m_\text{C}$ is the total mass of the SC. From this we can find the spin $S_\text{C}$ and the scalar mass quadrupole $Q_\text{C}\equiv\int\di^3\mb x\,\rho(\mb x)x^2\text{P}_2(\cos\theta)$ as,
\begin{align}
\label{SCQC}
  &S_\text{C}=\al m_\text{C}r_\text{B},
  &&Q_\text{C}=-6m_\text{C}r_\text{B}^2,
\end{align}
where $r_\text{B}\equiv 1/(\al\nu)$ is the Bohr radius. Assuming the BH with almost maximal initial spin, the cloud mass can reach $m_\text{C}=\al M$ when the production is saturated. Therefore we see that the cloud carries $\order{\al}$ of total mass of the GA but carries almost all the angular momentum and mass quadrupole of the GA when $\al\ll 1$.

\noindent{\bf GA-induced orbital precessions.} Now consider a GA-borne binary with eccentric orbit. The GA’s oblate energy distribution produces a deviation from Keplerian potential already at the Newtonian level. Consequently, the orbit will exhibit both apsidal and nodal precessions which we shall review below.

Let the two binary members have masses $m_1$ and $m_2$. We shall also use $m\equiv m_1+m_2$ and $\mu\equiv m_1m_2/m$. Let the first mass $m_1$ be a GA, consisting of a BH of mass $m_B$ and a saturated SC in $|211\ra$ of total mass $m_\text{C}\simeq\al m_1$, so that $m_1=m_\text{B}+m_\text{C}$. We assume that the GA is almost maximally spinning, with angular momentum $S_1=|\mb S_1|\simeq G m_1^2$. The mass quadrupole tensor of GA is given by $Q^{ij}= Q_1(\wh S_1^i\wh S_1^j-\de^{ij}/3)$ where $Q_1 = Q_\text{C}$ is the scalar quadrupole given in (\ref{SCQC}) and $\wh S_1^i\equiv S_1^i/S_1$. The second object $m_2$ will be taken as a point mass. It can possess finite spin but we will neglect it in most of discussions.

We focus on the early inspiral period of the binary evolution where the Post-Newtonian (PN) approximation is easily valid. Our signal, the quadrupole-induced precession, is a Newtonian effect. But it is also important to include higher PN effects that contribute orbital precessions, including the GR-induced apsidal precession at 1PN and the spin-orbit (SO) coupling at 1.5PN. The spin-spin (SS) coupling at 2PN is always negligible in our system. The GR precession can in principle be subtracted once we know the masses and orbital parameters, while SO- and SS-induced precessions depend on more free parameters, and thus should be regarded as contaminations of our signal.

The GA is also subject to tidal deformation and this can back react to the orbit and affects GW signals \cite{Cardoso:2020hca,Takahashi:2021eso,DeLuca:2021ite}. But this belongs to higher PN effects and we will neglect them in this Letter. The validity of treating our system as a well separated binary also puts a constraint on the orbital separation $a$. Following \cite{Baumann:2018vus} we take $a>4r_\text{B}$ for $|211\ra$ state, also shown in Fig.\;\ref{fig_alphaDist}. We don’t include mass ratio dependence in this constraint because we will never consider the case of $m_2\gg m_1$ in this Letter.

With above considerations in mind, we can model the GA as a rigid symmetric top, described by its mass $m_1$, spin $\mb S_1$, and quadrupole $Q_1$. We will adopt the famous EIH Lagrangian at 1PN to describe the barycenter motion of the two masses \cite{Einstein:1938yz}, adding appropriate terms to include rotational degrees of the GA. In particular, the interaction potential $V = V_\text{O} + V_\text{Q} + V_\text{SO}$ contains the point-mass term, the quadrupole term, and SO term, respectively,
\begin{align}
  V_\text{O}=&-\FR{Gm\mu}{r}+\text{1PN terms},\\
  \label{VQ}
  V_\text{Q}=&~\FR{Gm\mu}{2r^3}q_1(1-3\cos^2n_1),\\
  V_\text{SO}=&~\FR{2GL}{r^3}(m_1+\fr{3}{4}m_2)s_1\cos n_1,
\end{align} 
where $n_1$ is the angle between $\mb S_1$ and the orbital angular momentum $\mb L$. We further define $q_1\equiv Q_1/m_1$, $s_1\equiv S_1/m_1$, and $L\equiv|\mb L|$.

The computation of precession rates of various orbital elements from the above potential is straightforward, of which we will review systematically in a companion paper \cite{preclong}. As mentioned, there are both nodal precession (precession of $L$) and the apsidal precession (precession of the periapsis \emph{within} the orbital plane). We discuss them in turn.

First the nodal precession. Our system always has $L\gg S_1$, and so the angular momentum conservation dictates that $\mb L$ precesses only within a small cone of angular size $\sim S_1/L$. Therefore, the nodal precession only leaves small modulation of GW amplitudes (and phases) in generic situation. The only exception is when the line of sight overlaps with the small precession cone. In this case the nodal precession mixes with the orbital motion. But this is a rare situation. So the nodal precession is unimportant for most GW sources. For pulsar-BH binaries, however, the nodal precession can still be quite significant due to the high precision of pulsar timing measurements.

The most important effect is the apsidal precession. Let $\chi$ be the argument of periapsis in the orbital plane. As mentioned, the apsidal precession $\dot\chi$ has three contributions, from the GR precession at 1PN, mass quadrupole coupling $V_\text{Q}$, and the SO coupling $V_\text{SO}$, re-spectively:
\begin{align}
  \label{dotchi}
  \dot\chi=&~\dot\chi_\text{GR}+\dot\chi_\text{Q}+\dot\chi_\text{S},\\
  \dot\chi_\text{GR}=&~\FR{3Gm\omega}{a(1-e^2)},\\
  \label{dotchiQ}
  \dot\chi_\text{Q}=&~\FR{3\omega}{4a^2(1-e^2)^2}q_1(1-3\cos^2n_1),\\
  \dot\chi_\text{S}=&~\FR{-3\omega^2}{(1-e^2)^{3/2}}s_1\Big(1+\FR{m_1}{3m}\Big)\cos n_1,
\end{align}
where $a$ and $e$ are semi-major axis and the eccentricity of the orbit, and $\omega=\sqrt{Gm/a^3}$ is the orbital frequency. Let us compare their sizes. For maximally spinning and saturated GA in $|211\ra$, we have $s_1\simeq Gm_1$ and $q_1\simeq -6\al^3(Gm_1)^2$. Then,
\begin{align}
  &\FR{\dot\chi_\text{Q}}{\dot\chi_\text{GR}}\simeq \FR{3u_1}{2\al^3}\FR{Gm_1}{a(1-e^2)},\\
  &\FR{\dot\chi_\text{Q}}{\dot\chi_\text{S}}\simeq\FR{3u_1}{2\al^3(1+u_1/3)}\Big[\FR{Gm_1}{a(1-e^2)}\Big]^{1/2},
\end{align}
where $u_1\equiv m_1/m$. The factor $\al^{-3}$ from the large mass quadrupole is an enhancement while the factor $Gm_1/a<\al^2/4$ is always a suppression after imposing the perturbative condition $a>4r_\text{B}$. (One can impose a stronger condition $a(1-e)>4r_\text{B}$ but we will neglect the $(1-e)$ factor since we do not consider very large $e\simeq 1$.)

The GR precession can in principle be subtracted from the total $\dot\chi$ up to measurement errors once we know $a$, $e$, and $m$. On the other hand, $\dot\chi_\text{S}$ contaminates our signal by introducing new free parameter $s_1$. There is a similar contamination from the spin of $m_2$ which we assume to be no larger than that of $m_1$ and thus is suppressed here. We impose a constraint on a by requiring $|\dot\chi_\text{Q}|\gg |\dot\chi_\text{S}|$, to stay away from the SO-induced contamination, shown in Fig.\;\ref{fig_alphaDist}. The SS-induced precession is always smaller than SO-precession and thus can be safely neglected. Then, in the range of $|\dot\chi_\text{Q}|\gg |\dot\chi_\text{S}|$ and after subtracting the GR precession, the quadrupole term dominates the apsidal precession, leaving a distinct signal. Interestingly, the GR precession is always prograde while the quadrupole precession can be either prograde or retrograde.

The spin $\mb S_1$ also precesses due to the torque from quadrupole and SO couplings, and one may hope to probe this from the time variation of $\cos n_1=\wh{\mb L}\cdot\wh{\mb S}_1$. However, since $L$, $S$, and $\mb L +\mb S$ are all conserved in our system, it follows that $n_1$ is also a constant in time. Interestingly, $\dot\chi_\text{Q}$ remains constant even when $m_2$ carries nonzero quadrupole in which case $n_1$ is no longer a constant. We will elaborate more on this in a companion paper \cite{preclong}.

In summary, by staying in the range of $|\dot\chi_\text{Q}|\gg |\dot\chi_\text{S}|$, only $\dot\chi_\text{GR}$ and $\dot\chi_\text{Q}$ contribute the apsidal precession, and our signal, the quadrupole induced apsidal precession, depends on a single constant parameter which we define as $q_\text{eff}$ for later convenience:
\bge
  q_\text{eff}\equiv q_1(1-3\cos^2n_1).
\ede

\noindent{\bf GWs from GA-BH Binaries.} The GWs radiated from a binary essentially trace the binary’s orbital motion. Thus one could hope to see the orbital precession from the GW waveform. But this is challenging since we need to see the GW from early inspiral phase rather than close to the merger. Early inspiral is required both to have nonzero eccentricity and also to satisfy the perturbative condition (i.e., the second body be well outside the cloud). Requiring $a>4r_\text{B}$, the GW frequency (of the 2nd harmonic) $f_2=\pi^{-1}\sqrt{Gm/a^3}<\pi^{-1}\sqrt{Gm/(4r_\text{B})^3}\sim0.14\text{Hz}(\al/0.05)^3(10M_\odot/m_1)$, assuming $m_1=m_2$. Thus we see that for $10M_\odot$ GA the signal is already in the window of LISA. The similar signal from massive BH binaries ($m>10^4M_\odot$) or extreme-mass-ratio inspirals (typically a supermassive BH with a stellar-mass BH) will be at even lower frequencies, challenging to resolve individually even by space GW telescope. Therefore we will focus on stellar-mass binaries in this section and leave more general possibilities to a future study.

The GW amplitudes from the binary in the transverse-traceless (TT) gauge is computed by $h^{ij}=(2G/d)\Lambda^{ij,k\ell}\ddot M^{k\ell}$ where $\Lambda^{ij,k\ell}$ is the projector to TT components, $M^{ij}=\mu r^ir^j$ is the second mass moment of the binary, and $d$ is the source distance. In our case $M^{ij}$ depends on $a$, $e$, the true anomaly $\psi(t)$, and the three Euler angles describing the orbital orientation. These angles can be chosen to be the argument of periapsis $\chi$, the inclination $I$ and the longitude of ascending node $\Omega$, defined with respect to a reference plane (e.g., the plane perpendicular to the line of sight). In a GA-borne binary, all three angles could precess and this precession could enter the GW signal. But as discussed above, in generic situations, the nodal precession is tiny, and only the apsidal precession $\dot\chi$ could potentially be observable.

To identify the signature of $\dot\chi$ in GW signals, we go to the orbital plane, in which the position vector $r_i=r(\cos\ob\psi,\sin\ob\psi)$, with $r=a(1-e^2)/(1+e\cos\psi)$, and $\ob\psi=\psi+\chi$. (For low inclination system $I\simeq 0$, one should also include the motion of ascending node so that $\ob\psi=\psi+\chi+\Omega$. However, once again, we consider such systems rare.) The important point is that the radial motion $r(\psi)$ depends only on the true anomaly which is a periodic function of time with frequency $\psi$, while the angular motion also includes the apsidal precession, so that the angle $\ob\psi$ is a periodic function with frequency $\omega+\dot\chi$.

\begin{figure}
\centering
\includegraphics[width=0.32\textwidth]{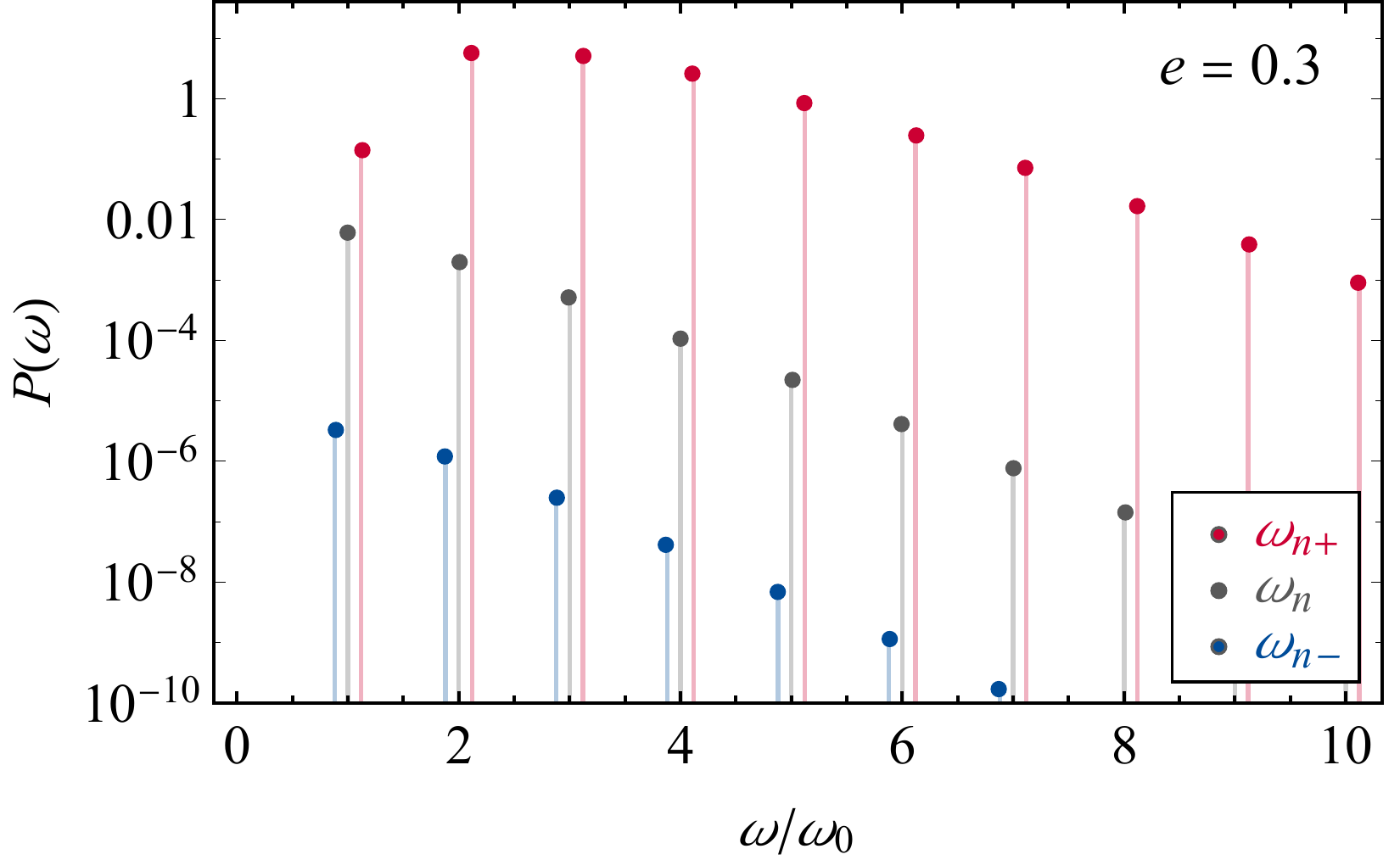}
\caption{The GW power spectrum of an eccentric binary with apsidal precession. The split in each triplet is exaggerated.}
\label{fig_spect}
\end{figure}

To understand intuitively how to observe $\dot\chi$ through GWs, we note that $\psi(t)$ executes non-uniform motion for eccentric orbit (Kepler’s 2nd law), resulting overtones in GW spectrum with frequency $\omega_n=n\omega$ $(n=1,2,\cdots)$. We can assume $\dot\chi$ to be almost constant up to slight orbital decay. Then, each of $M^{ij}$’s harmonic components splits into a triplet with frequencies $(\omega_n,\omega_{n\pm})\equiv(n\omega,n\omega\pm2\dot\chi)$. It turns out that the $\omega_{n+}$ component almost always dominates the GW power in a triplet, as shown in Fig.\;\ref{fig_spect}. Therefore we can retain the $\omega_{n+}$ component only. Then the effect of apsidal precession enters through the \emph{anharmonic overtones}, namely, $\omega_{n+}/\omega_{m+}$ is not integer for any $n\neq m$. Of course the visibility of this non-integer ratio depends on the frequency resolution, which requires $\dot\chi_\text{Q}>T_O^{-1}$ with $T_O$ the total observation time. Using (\ref{dotchiQ}) this translates very roughly to $q_1/a^2>(\omega T_O)^{-1}$. Further using $q_1\sim\al r_\text{B}^2$ and the perturbativity condition $a\gtrsim r_\text{B}$, we have $q_1/a^2<\al$. On the other hand, taking LISA as an example, $\omega\sim$0.01Hz and $T_O\sim$\;yr, so $\omega T_O \sim10^5$. So there is a finite range $10^{-5}<q_1/a^2<\al$ in which we can hope to see $\dot\chi_\text{Q}$ through GWs.

We note that this effect is absent for circular orbit where $r(\psi)$ is a constant, and $\ob\psi=(\omega+\dot\chi)t$ is essentially a uniform function of time. So a nonzero $\dot\chi$ amounts to a constant shift in $\omega$ and thus is degenerate with other orbital parameters. Of course this degeneracy is broken by PN effects including chirping. But we expect less significant effect from the quadrupole in this case. Below we will show that increasing $e$ reduces the error in $q_\text{eff}$.

\begin{figure}
\centering
\includegraphics[width=0.49\textwidth]{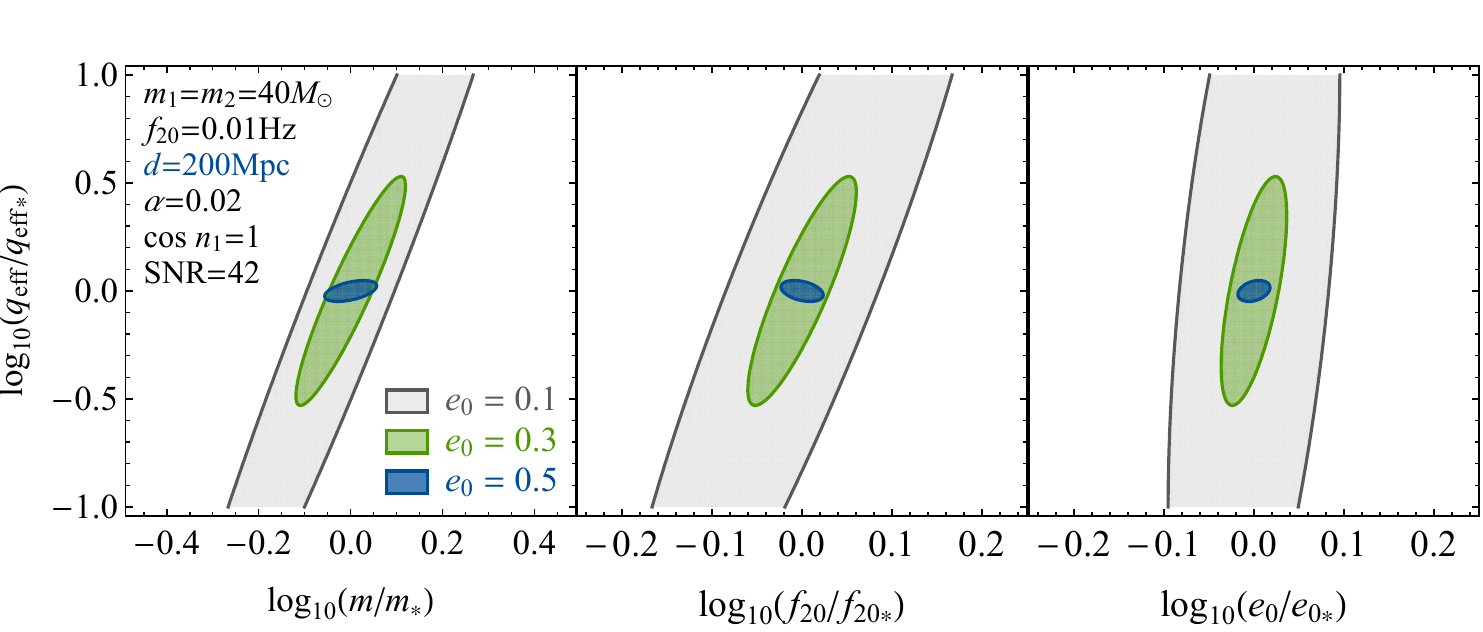} 
\caption{The $1\si$ error contours of $q_\text{eff}$ with the total mass $m$, the GW frequency of 2nd harmonic $f_{20}$, and the eccentricity $e_0$. We choose N2A5 configuration of LISA and 4yr of total observation time. The choice of parameters for this plot is a bit beyond the bound $a>4r_\text{B}$ so the result is only indicative.}
\label{fig_GWerr}
\end{figure}

To assess more quantitatively the observability of the quadrupole-induced apsidal precession through GW, we perform a simplified Fisher analysis for a binary of $(40 +40)M_\odot$. A more complete analysis will be presented in \cite{preclong}. Here we consider a 4-parameter set that directly appears in $\dot\chi$, including $q_\text{eff}$, the total mass $m$, the GW frequency $f_{20}$ of 2nd harmonic and eccentricity $e_0$ at the starting time of observation. This parameter set is useful to address potential degeneracy between $\dot\chi_\text{GR}$ and $\dot\chi_\text{Q}$. We apply a time-domain formula for Fisher matrix \cite{Barack:2003fp} and assume N2A5 configuration of LISA noise curve \cite{Klein:2015hvg} with 4yr of total observation time. For our purpose it is enough to use the Newtonian orbit augmented by apsidal precession (8). We apply Peters’s equation \cite{Peters:1964zz} to account for orbital decay but treat Euler angles $(I,\Omega)$ as constant. In Fig.\;\ref{fig_GWerr} we show the $1\si$ contours for $e_0 = (0.1, 0.3, 0.5)$, with all other parameters indicated in the Figure. The signal-to-noise ratios (SNR) for three choices of e are normalized to that of $e = 0.5$ at $d = 200$Mpc, which is roughly SNR\;$\simeq42$. One can see that the mass quadrupole $q_\text{eff}$ can be clearly identified for $e = 0.3$ and well measured for $e = 0.5$, while large degeneracy and errors appear for $e = 0.1$, just as expected.

\noindent{\bf Cloud-induced precession in pulsar-GA binaries.} The effect of the superradiance cloud can also be searched for in pulsar-BH binaries. Compared to GWs, the very precise pulsar timing data usually allow more direct measurement of orbital elements. The apsidal precession rate $\dot\chi$ becomes almost a direct observable readily to be extracted from the time delay data.

In pulsar binaries the 5 ``Keplerian'' parameters describing the geometry and kinetics of the orbit can often be very precisely measured, while in well situated systems one can also measure several post-Keplerian (PK) parameters with good precision \cite{Stairs:2003eg}. In GR, all PK parameters can be calculated from Keplerian parameters and the two masses. Therefore measuring two PK parameters can determine the two masses, conventionally denoted by $m_p$ for the pulsar and $m_c$ for the companion. (With our notation, $m_1=m_c$ and $m_2=m_p$.) Measuring more than 2 PK parameters then serves as consistency check. It is often the case that the apsidal precession rate $\dot\chi$, the Einstein parameter $\ga$, and the orbital decay $\dot P_b$ can be measured to high precision. For highly inclined system the Shapiro delay parameters $r$ and $s$ can also be measured.

\begin{figure}
\centering
\includegraphics[width=0.26\textwidth]{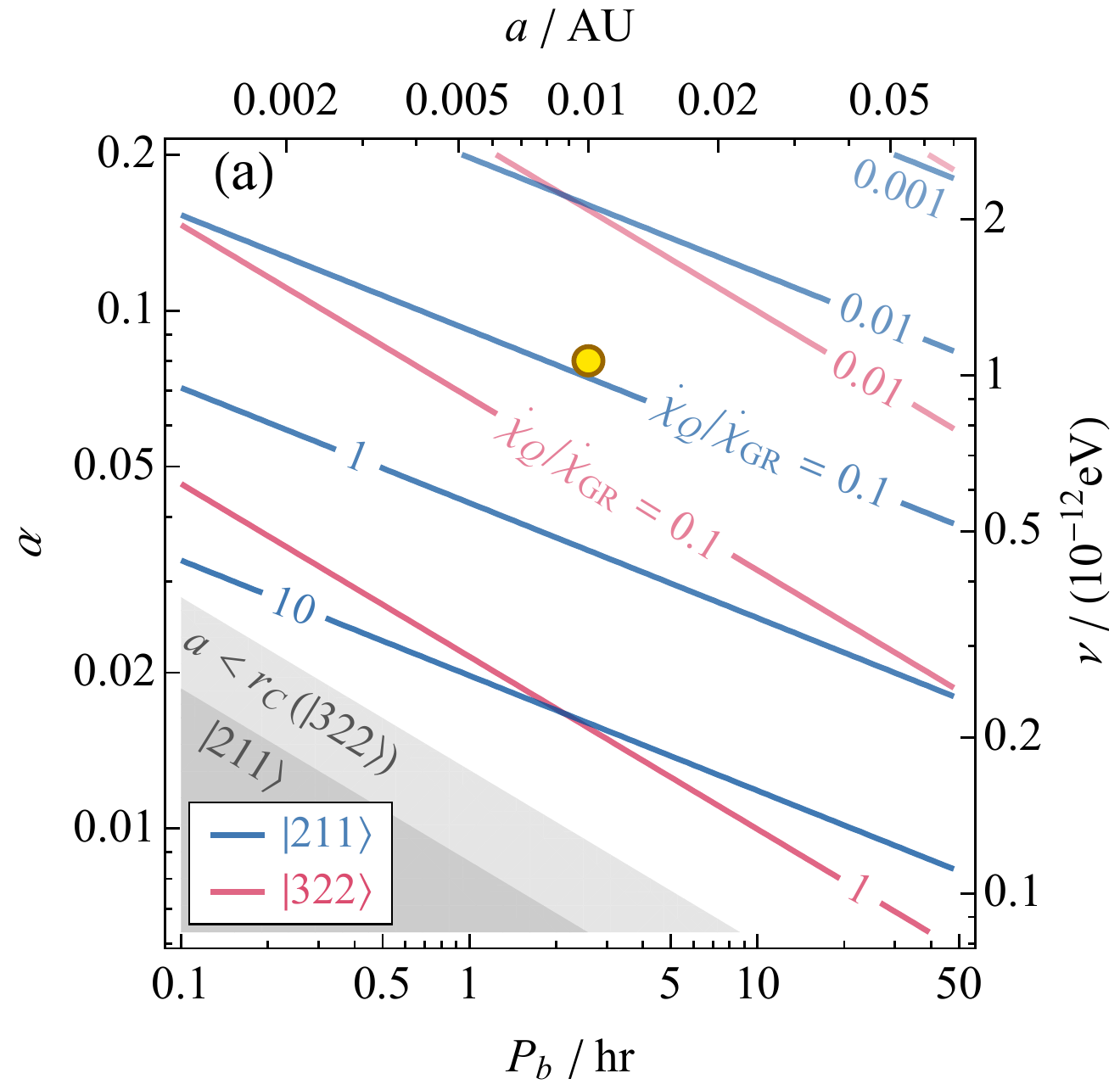}~
\includegraphics[width=0.22\textwidth]{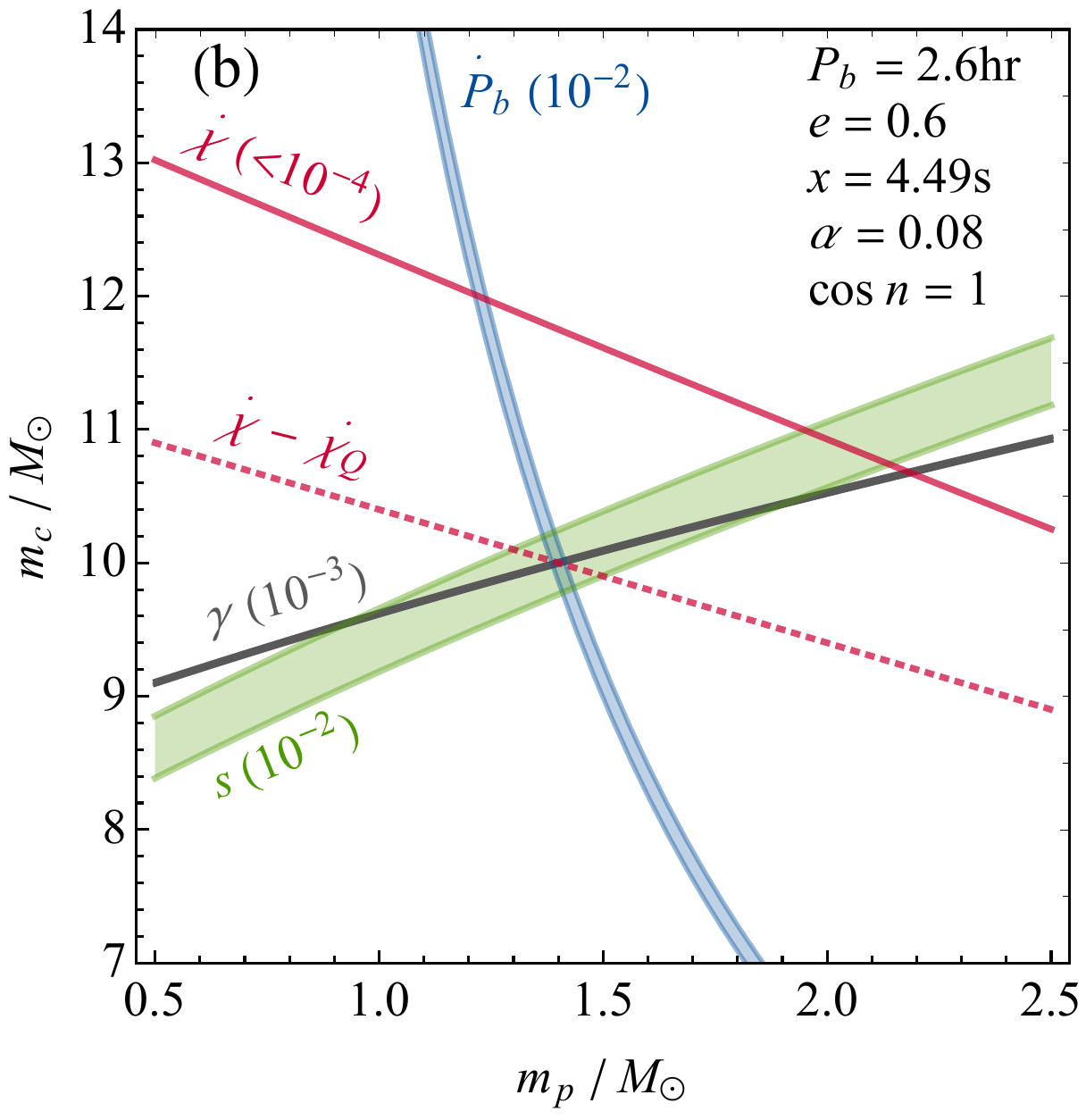}
\caption{(a) The comparison between the cloud-induced apsidal precession rate $\dot\chi_\text{Q}$ and the GR induced precession rate $\dot\chi_\text{GR}$ for a range of $\al$ and orbital period $P_b$. In this plot we take $m_p=1.4M_\odot$, $m_c=10M_\odot$, $e = 0.6$, and assume the cloud mass is saturated. (b) The PK parameters fit in $(m_p,m_c)$ plane for a pulsar-GA binary marked by the yellow dot in (a). The error bands, shown in parentheses, are taken with the same order of magnitude from known pulsar binaries such as PSR B1913+16 and PSR B1534+12 \cite{Stairs:2003eg}. The solid and dashed magenta curves show the $\dot\chi$ without and with subtracting the quadrupole contribution, respectively.}
\label{fig_pulsar}
\end{figure}

In our case, however, the orbital evolution depends on further parameters including $q_1$ and $n_1$ which correct PK parameters. Most of them are small corrections of order $q_1/a^2\ll 1$. But there are two post-Keplerian parameters whose quadrupole correction is not suppressed by $q_1/a^2$. The most important one is of course $\dot\chi$ given in (\ref{dotchi}) (normally denoted by $\omega$ in pulsar binary literature). In Fig.\;\ref{fig_pulsar}(a) we show the comparison between $\dot\chi_\text{Q}$ and $\dot\chi_\text{GR}$ for a range of orbital period $P_b$ and $\al$. We assume $m_c$ is a GA saturated in either $|211\ra$ or $|322\ra$. We see that the correction from $\dot\chi_\text{Q}$ is actually huge compared with measurement precision, which is easily better than $10^{-3}$. In Fig.\;\ref{fig_pulsar}(a) we also put a yellow spot and show an imagined PK parameter fit in the $(m_p,m_c)$ plane for this data point as is usually done for pulsar binaries. Consistency requires that all curves meet at one point up to errors. However, for the parameter we are choosing ($\dot\chi_\text{Q}/\dot\chi_\text{GR}\simeq 0.1$), the failure of the consistency check is already dramatic, showing that there can be a good chance to measure $\dot\chi_\text{Q}$ in this way.

The other PK parameter receiving significant correction is $\dot x$ where $x\equiv a\sin I$. In ordinary system $\dot x$ comes mainly from the GW-induced orbital decay $\dot a$. This is a tiny effect and $\dot x$ has not yet been measured in known systems. In our system, however, the inclination $\sin I$ becomes time dependent due to quadrupole coupling (\ref{VQ}). One can show that $\dot x$ has the same time scale as $\dot\chi_\text{Q}$ and so is easily within the observation range. The total variation of $x$ remains small as mentioned before, $\Delta x/x\sim\order{S/L}$, due to the angular momentum conservation. But for the system plotted in Fig.\;\ref{fig_pulsar}(b), $S/L\simeq 0.06$ and so can easily be larger than observation precision of $x$, making $\dot x$ potentially observable. We will present a more systematic study of this effect in a future work.

\noindent{\bf Discussions.} Fast rotating BHs are natural factories and habitat for ultralight bosons. The resulting GAs are ideal labs for probing these otherwise invisible particles. In this Letter we showed that the large mass quadrupole of a GA can excite observable apsidal precession for eccentric binaries. The effect does not require resonant transitions and thus extends the reach of gravitational collider physics. It will be interesting to consider more such non-resonant effects in similar systems, such as a binary of two GAs, and also a ``gravitational molecule'' with one SC surrounding two BHs \cite{Ikeda:2020xvt,Liu:2021llm}. We leave these directions for future works.

\begin{acknowledgements}
We thank Jiandong Zhang for helpful discussions. We also thank Horng Sheng Chia, Jing Ren, and Xi Tong for useful comments on the manuscript. ZZX is supported by Tsinghua University Initiative Scientific Research Program. 
\end{acknowledgements}


\begin{thebibliography}{59}%
\makeatletter
\providecommand \@ifxundefined [1]{%
 \@ifx{#1\undefined}
}%
\providecommand \@ifnum [1]{%
 \ifnum #1\expandafter \@firstoftwo
 \else \expandafter \@secondoftwo
 \fi
}%
\providecommand \@ifx [1]{%
 \ifx #1\expandafter \@firstoftwo
 \else \expandafter \@secondoftwo
 \fi
}%
\providecommand \natexlab [1]{#1}%
\providecommand \enquote  [1]{``#1''}%
\providecommand \bibnamefont  [1]{#1}%
\providecommand \bibfnamefont [1]{#1}%
\providecommand \citenamefont [1]{#1}%
\providecommand \href@noop [0]{\@secondoftwo}%
\providecommand \href [0]{\begingroup \@sanitize@url \@href}%
\providecommand \@href[1]{\@@startlink{#1}\@@href}%
\providecommand \@@href[1]{\endgroup#1\@@endlink}%
\providecommand \@sanitize@url [0]{\catcode `\\12\catcode `\$12\catcode
  `\&12\catcode `\#12\catcode `\^12\catcode `\_12\catcode `\%12\relax}%
\providecommand \@@startlink[1]{}%
\providecommand \@@endlink[0]{}%
\providecommand \url  [0]{\begingroup\@sanitize@url \@url }%
\providecommand \@url [1]{\endgroup\@href {#1}{\urlprefix }}%
\providecommand \urlprefix  [0]{URL }%
\providecommand \Eprint [0]{\href }%
\providecommand \doibase [0]{http://dx.doi.org/}%
\providecommand \selectlanguage [0]{\@gobble}%
\providecommand \bibinfo  [0]{\@secondoftwo}%
\providecommand \bibfield  [0]{\@secondoftwo}%
\providecommand \translation [1]{[#1]}%
\providecommand \BibitemOpen [0]{}%
\providecommand \bibitemStop [0]{}%
\providecommand \bibitemNoStop [0]{.\EOS\space}%
\providecommand \EOS [0]{\spacefactor3000\relax}%
\providecommand \BibitemShut  [1]{\csname bibitem#1\endcsname}%
\let\auto@bib@innerbib\@empty
\bibitem [{\citenamefont {Abbott}\ \emph {et~al.}(2016)\citenamefont {Abbott}
  \emph {et~al.}}]{LIGOScientific:2016aoc}%
  \BibitemOpen
  \bibfield  {author} {\bibinfo {author} {\bibfnamefont {B.~P.}\ \bibnamefont
  {Abbott}} \emph {et~al.} (\bibinfo {collaboration} {LIGO Scientific,
  Virgo}),\ }\href {\doibase 10.1103/PhysRevLett.116.061102} {\bibfield
  {journal} {\bibinfo  {journal} {Phys. Rev. Lett.}\ }\textbf {\bibinfo
  {volume} {116}},\ \bibinfo {pages} {061102} (\bibinfo {year} {2016})},\
  \Eprint {http://arxiv.org/abs/1602.03837} {arXiv:1602.03837 [gr-qc]}
  \BibitemShut {NoStop}%
\bibitem [{\citenamefont {Perkins}\ \emph {et~al.}(2021)\citenamefont
  {Perkins}, \citenamefont {Yunes},\ and\ \citenamefont
  {Berti}}]{Perkins:2020tra}%
  \BibitemOpen
  \bibfield  {author} {\bibinfo {author} {\bibfnamefont {S.~E.}\ \bibnamefont
  {Perkins}}, \bibinfo {author} {\bibfnamefont {N.}~\bibnamefont {Yunes}}, \
  and\ \bibinfo {author} {\bibfnamefont {E.}~\bibnamefont {Berti}},\ }\href
  {\doibase 10.1103/PhysRevD.103.044024} {\bibfield  {journal} {\bibinfo
  {journal} {Phys. Rev. D}\ }\textbf {\bibinfo {volume} {103}},\ \bibinfo
  {pages} {044024} (\bibinfo {year} {2021})},\ \Eprint
  {http://arxiv.org/abs/2010.09010} {arXiv:2010.09010 [gr-qc]} \BibitemShut
  {NoStop}%
\bibitem [{\citenamefont {Essig}\ \emph {et~al.}(2013)\citenamefont {Essig}
  \emph {et~al.}}]{Essig:2013lka}%
  \BibitemOpen
  \bibfield  {author} {\bibinfo {author} {\bibfnamefont {R.}~\bibnamefont
  {Essig}} \emph {et~al.},\ }in\ \href@noop {} {\emph {\bibinfo {booktitle}
  {{Community Summer Study 2013}: {Snowmass on the Mississippi}}}}\ (\bibinfo
  {year} {2013})\ \Eprint {http://arxiv.org/abs/1311.0029} {arXiv:1311.0029
  [hep-ph]} \BibitemShut {NoStop}%
\bibitem [{\citenamefont {Zeldovich}(1971)}]{Zeldovich1971}%
  \BibitemOpen
  \bibfield  {author} {\bibinfo {author} {\bibfnamefont {Y.~B.}\ \bibnamefont
  {Zeldovich}},\ }\href@noop {} {\bibfield  {journal} {\bibinfo  {journal}
  {JETP Lett.}\ }\textbf {\bibinfo {volume} {14}},\ \bibinfo {pages} {180}
  (\bibinfo {year} {1971})}\BibitemShut {NoStop}%
\bibitem [{\citenamefont {Starobinsky}(1973)}]{Starobinsky:1973aij}%
  \BibitemOpen
  \bibfield  {author} {\bibinfo {author} {\bibfnamefont {A.~A.}\ \bibnamefont
  {Starobinsky}},\ }\href@noop {} {\bibfield  {journal} {\bibinfo  {journal}
  {Sov. Phys. JETP}\ }\textbf {\bibinfo {volume} {37}},\ \bibinfo {pages} {28}
  (\bibinfo {year} {1973})}\BibitemShut {NoStop}%
\bibitem [{\citenamefont {Brito}\ \emph {et~al.}(2015)\citenamefont {Brito},
  \citenamefont {Cardoso},\ and\ \citenamefont {Pani}}]{Brito:2015oca}%
  \BibitemOpen
  \bibfield  {author} {\bibinfo {author} {\bibfnamefont {R.}~\bibnamefont
  {Brito}}, \bibinfo {author} {\bibfnamefont {V.}~\bibnamefont {Cardoso}}, \
  and\ \bibinfo {author} {\bibfnamefont {P.}~\bibnamefont {Pani}},\ }\href
  {\doibase 10.1007/978-3-319-19000-6} {\bibfield  {journal} {\bibinfo
  {journal} {Lect. Notes Phys.}\ }\textbf {\bibinfo {volume} {906}},\ \bibinfo
  {pages} {pp.1} (\bibinfo {year} {2015})},\ \Eprint
  {http://arxiv.org/abs/1501.06570} {arXiv:1501.06570 [gr-qc]} \BibitemShut
  {NoStop}%
\bibitem [{\citenamefont {Arvanitaki}\ and\ \citenamefont
  {Dubovsky}(2011)}]{Arvanitaki:2010sy}%
  \BibitemOpen
  \bibfield  {author} {\bibinfo {author} {\bibfnamefont {A.}~\bibnamefont
  {Arvanitaki}}\ and\ \bibinfo {author} {\bibfnamefont {S.}~\bibnamefont
  {Dubovsky}},\ }\href {\doibase 10.1103/PhysRevD.83.044026} {\bibfield
  {journal} {\bibinfo  {journal} {Phys. Rev. D}\ }\textbf {\bibinfo {volume}
  {83}},\ \bibinfo {pages} {044026} (\bibinfo {year} {2011})},\ \Eprint
  {http://arxiv.org/abs/1004.3558} {arXiv:1004.3558 [hep-th]} \BibitemShut
  {NoStop}%
\bibitem [{\citenamefont {Cardoso}\ \emph {et~al.}(2011)\citenamefont
  {Cardoso}, \citenamefont {Chakrabarti}, \citenamefont {Pani}, \citenamefont
  {Berti},\ and\ \citenamefont {Gualtieri}}]{Cardoso:2011xi}%
  \BibitemOpen
  \bibfield  {author} {\bibinfo {author} {\bibfnamefont {V.}~\bibnamefont
  {Cardoso}}, \bibinfo {author} {\bibfnamefont {S.}~\bibnamefont
  {Chakrabarti}}, \bibinfo {author} {\bibfnamefont {P.}~\bibnamefont {Pani}},
  \bibinfo {author} {\bibfnamefont {E.}~\bibnamefont {Berti}}, \ and\ \bibinfo
  {author} {\bibfnamefont {L.}~\bibnamefont {Gualtieri}},\ }\href {\doibase
  10.1103/PhysRevLett.107.241101} {\bibfield  {journal} {\bibinfo  {journal}
  {Phys. Rev. Lett.}\ }\textbf {\bibinfo {volume} {107}},\ \bibinfo {pages}
  {241101} (\bibinfo {year} {2011})},\ \Eprint {http://arxiv.org/abs/1109.6021}
  {arXiv:1109.6021 [gr-qc]} \BibitemShut {NoStop}%
\bibitem [{\citenamefont {Yoshino}\ and\ \citenamefont
  {Kodama}(2012)}]{Yoshino:2012kn}%
  \BibitemOpen
  \bibfield  {author} {\bibinfo {author} {\bibfnamefont {H.}~\bibnamefont
  {Yoshino}}\ and\ \bibinfo {author} {\bibfnamefont {H.}~\bibnamefont
  {Kodama}},\ }\href {\doibase 10.1143/PTP.128.153} {\bibfield  {journal}
  {\bibinfo  {journal} {Prog. Theor. Phys.}\ }\textbf {\bibinfo {volume}
  {128}},\ \bibinfo {pages} {153} (\bibinfo {year} {2012})},\ \Eprint
  {http://arxiv.org/abs/1203.5070} {arXiv:1203.5070 [gr-qc]} \BibitemShut
  {NoStop}%
\bibitem [{\citenamefont {Arvanitaki}\ \emph {et~al.}(2015)\citenamefont
  {Arvanitaki}, \citenamefont {Baryakhtar},\ and\ \citenamefont
  {Huang}}]{Arvanitaki:2014wva}%
  \BibitemOpen
  \bibfield  {author} {\bibinfo {author} {\bibfnamefont {A.}~\bibnamefont
  {Arvanitaki}}, \bibinfo {author} {\bibfnamefont {M.}~\bibnamefont
  {Baryakhtar}}, \ and\ \bibinfo {author} {\bibfnamefont {X.}~\bibnamefont
  {Huang}},\ }\href {\doibase 10.1103/PhysRevD.91.084011} {\bibfield  {journal}
  {\bibinfo  {journal} {Phys. Rev. D}\ }\textbf {\bibinfo {volume} {91}},\
  \bibinfo {pages} {084011} (\bibinfo {year} {2015})},\ \Eprint
  {http://arxiv.org/abs/1411.2263} {arXiv:1411.2263 [hep-ph]} \BibitemShut
  {NoStop}%
\bibitem [{\citenamefont {Abbott}\ \emph {et~al.}(2019)\citenamefont {Abbott}
  \emph {et~al.}}]{LIGOScientific:2018mvr}%
  \BibitemOpen
  \bibfield  {author} {\bibinfo {author} {\bibfnamefont {B.~P.}\ \bibnamefont
  {Abbott}} \emph {et~al.} (\bibinfo {collaboration} {LIGO Scientific,
  Virgo}),\ }\href {\doibase 10.1103/PhysRevX.9.031040} {\bibfield  {journal}
  {\bibinfo  {journal} {Phys. Rev. X}\ }\textbf {\bibinfo {volume} {9}},\
  \bibinfo {pages} {031040} (\bibinfo {year} {2019})},\ \Eprint
  {http://arxiv.org/abs/1811.12907} {arXiv:1811.12907 [astro-ph.HE]}
  \BibitemShut {NoStop}%
\bibitem [{\citenamefont {Abbott}\ \emph {et~al.}(2021)\citenamefont {Abbott}
  \emph {et~al.}}]{LIGOScientific:2020ibl}%
  \BibitemOpen
  \bibfield  {author} {\bibinfo {author} {\bibfnamefont {R.}~\bibnamefont
  {Abbott}} \emph {et~al.} (\bibinfo {collaboration} {LIGO Scientific,
  Virgo}),\ }\href {\doibase 10.1103/PhysRevX.11.021053} {\bibfield  {journal}
  {\bibinfo  {journal} {Phys. Rev. X}\ }\textbf {\bibinfo {volume} {11}},\
  \bibinfo {pages} {021053} (\bibinfo {year} {2021})},\ \Eprint
  {http://arxiv.org/abs/2010.14527} {arXiv:2010.14527 [gr-qc]} \BibitemShut
  {NoStop}%
\bibitem [{\citenamefont {Baumann}\ \emph
  {et~al.}(2019{\natexlab{a}})\citenamefont {Baumann}, \citenamefont {Chia},\
  and\ \citenamefont {Porto}}]{Baumann:2018vus}%
  \BibitemOpen
  \bibfield  {author} {\bibinfo {author} {\bibfnamefont {D.}~\bibnamefont
  {Baumann}}, \bibinfo {author} {\bibfnamefont {H.~S.}\ \bibnamefont {Chia}}, \
  and\ \bibinfo {author} {\bibfnamefont {R.~A.}\ \bibnamefont {Porto}},\ }\href
  {\doibase 10.1103/PhysRevD.99.044001} {\bibfield  {journal} {\bibinfo
  {journal} {Phys. Rev. D}\ }\textbf {\bibinfo {volume} {99}},\ \bibinfo
  {pages} {044001} (\bibinfo {year} {2019}{\natexlab{a}})},\ \Eprint
  {http://arxiv.org/abs/1804.03208} {arXiv:1804.03208 [gr-qc]} \BibitemShut
  {NoStop}%
\bibitem [{\citenamefont {Baumann}\ \emph
  {et~al.}(2019{\natexlab{b}})\citenamefont {Baumann}, \citenamefont {Chia},
  \citenamefont {Stout},\ and\ \citenamefont {ter Haar}}]{Baumann:2019eav}%
  \BibitemOpen
  \bibfield  {author} {\bibinfo {author} {\bibfnamefont {D.}~\bibnamefont
  {Baumann}}, \bibinfo {author} {\bibfnamefont {H.~S.}\ \bibnamefont {Chia}},
  \bibinfo {author} {\bibfnamefont {J.}~\bibnamefont {Stout}}, \ and\ \bibinfo
  {author} {\bibfnamefont {L.}~\bibnamefont {ter Haar}},\ }\href {\doibase
  10.1088/1475-7516/2019/12/006} {\bibfield  {journal} {\bibinfo  {journal}
  {JCAP}\ }\textbf {\bibinfo {volume} {12}},\ \bibinfo {pages} {006} (\bibinfo
  {year} {2019}{\natexlab{b}})},\ \Eprint {http://arxiv.org/abs/1908.10370}
  {arXiv:1908.10370 [gr-qc]} \BibitemShut {NoStop}%
\bibitem [{\citenamefont {Baumann}\ \emph {et~al.}(2020)\citenamefont
  {Baumann}, \citenamefont {Chia}, \citenamefont {Porto},\ and\ \citenamefont
  {Stout}}]{Baumann:2019ztm}%
  \BibitemOpen
  \bibfield  {author} {\bibinfo {author} {\bibfnamefont {D.}~\bibnamefont
  {Baumann}}, \bibinfo {author} {\bibfnamefont {H.~S.}\ \bibnamefont {Chia}},
  \bibinfo {author} {\bibfnamefont {R.~A.}\ \bibnamefont {Porto}}, \ and\
  \bibinfo {author} {\bibfnamefont {J.}~\bibnamefont {Stout}},\ }\href
  {\doibase 10.1103/PhysRevD.101.083019} {\bibfield  {journal} {\bibinfo
  {journal} {Phys. Rev. D}\ }\textbf {\bibinfo {volume} {101}},\ \bibinfo
  {pages} {083019} (\bibinfo {year} {2020})},\ \Eprint
  {http://arxiv.org/abs/1912.04932} {arXiv:1912.04932 [gr-qc]} \BibitemShut
  {NoStop}%
\bibitem [{\citenamefont {Ding}\ \emph {et~al.}(2021)\citenamefont {Ding},
  \citenamefont {Tong},\ and\ \citenamefont {Wang}}]{Ding:2020bnl}%
  \BibitemOpen
  \bibfield  {author} {\bibinfo {author} {\bibfnamefont {Q.}~\bibnamefont
  {Ding}}, \bibinfo {author} {\bibfnamefont {X.}~\bibnamefont {Tong}}, \ and\
  \bibinfo {author} {\bibfnamefont {Y.}~\bibnamefont {Wang}},\ }\href {\doibase
  10.3847/1538-4357/abd803} {\bibfield  {journal} {\bibinfo  {journal}
  {Astrophys. J.}\ }\textbf {\bibinfo {volume} {908}},\ \bibinfo {pages} {78}
  (\bibinfo {year} {2021})},\ \Eprint {http://arxiv.org/abs/2009.11106}
  {arXiv:2009.11106 [astro-ph.HE]} \BibitemShut {NoStop}%
\bibitem [{\citenamefont {Tong}\ \emph {et~al.}(2021)\citenamefont {Tong},
  \citenamefont {Wang},\ and\ \citenamefont {Zhu}}]{Tong:2021whq}%
  \BibitemOpen
  \bibfield  {author} {\bibinfo {author} {\bibfnamefont {X.}~\bibnamefont
  {Tong}}, \bibinfo {author} {\bibfnamefont {Y.}~\bibnamefont {Wang}}, \ and\
  \bibinfo {author} {\bibfnamefont {H.-Y.}\ \bibnamefont {Zhu}},\ }\href@noop
  {} {\  (\bibinfo {year} {2021})},\ \Eprint {http://arxiv.org/abs/2106.13484}
  {arXiv:2106.13484 [astro-ph.HE]} \BibitemShut {NoStop}%
\bibitem [{\citenamefont {Laarakkers}\ and\ \citenamefont
  {Poisson}(1999)}]{Laarakkers:1997hb}%
  \BibitemOpen
  \bibfield  {author} {\bibinfo {author} {\bibfnamefont {W.~G.}\ \bibnamefont
  {Laarakkers}}\ and\ \bibinfo {author} {\bibfnamefont {E.}~\bibnamefont
  {Poisson}},\ }\href {\doibase 10.1086/306732} {\bibfield  {journal} {\bibinfo
   {journal} {Astrophys. J.}\ }\textbf {\bibinfo {volume} {512}},\ \bibinfo
  {pages} {282} (\bibinfo {year} {1999})},\ \Eprint
  {http://arxiv.org/abs/gr-qc/9709033} {arXiv:gr-qc/9709033} \BibitemShut
  {NoStop}%
\bibitem [{\citenamefont {Pappas}\ and\ \citenamefont
  {Apostolatos}(2012)}]{Pappas:2012ns}%
  \BibitemOpen
  \bibfield  {author} {\bibinfo {author} {\bibfnamefont {G.}~\bibnamefont
  {Pappas}}\ and\ \bibinfo {author} {\bibfnamefont {T.~A.}\ \bibnamefont
  {Apostolatos}},\ }\href {\doibase 10.1103/PhysRevLett.108.231104} {\bibfield
  {journal} {\bibinfo  {journal} {Phys. Rev. Lett.}\ }\textbf {\bibinfo
  {volume} {108}},\ \bibinfo {pages} {231104} (\bibinfo {year} {2012})},\
  \Eprint {http://arxiv.org/abs/1201.6067} {arXiv:1201.6067 [gr-qc]}
  \BibitemShut {NoStop}%
\bibitem [{\citenamefont {Samsing}\ \emph {et~al.}(2014)\citenamefont
  {Samsing}, \citenamefont {MacLeod},\ and\ \citenamefont
  {Ramirez-Ruiz}}]{Samsing:2013kua}%
  \BibitemOpen
  \bibfield  {author} {\bibinfo {author} {\bibfnamefont {J.}~\bibnamefont
  {Samsing}}, \bibinfo {author} {\bibfnamefont {M.}~\bibnamefont {MacLeod}}, \
  and\ \bibinfo {author} {\bibfnamefont {E.}~\bibnamefont {Ramirez-Ruiz}},\
  }\href {\doibase 10.1088/0004-637X/784/1/71} {\bibfield  {journal} {\bibinfo
  {journal} {Astrophys. J.}\ }\textbf {\bibinfo {volume} {784}},\ \bibinfo
  {pages} {71} (\bibinfo {year} {2014})},\ \Eprint
  {http://arxiv.org/abs/1308.2964} {arXiv:1308.2964 [astro-ph.HE]} \BibitemShut
  {NoStop}%
\bibitem [{\citenamefont {Samsing}(2018)}]{Samsing:2017xmd}%
  \BibitemOpen
  \bibfield  {author} {\bibinfo {author} {\bibfnamefont {J.}~\bibnamefont
  {Samsing}},\ }\href {\doibase 10.1103/PhysRevD.97.103014} {\bibfield
  {journal} {\bibinfo  {journal} {Phys. Rev. D}\ }\textbf {\bibinfo {volume}
  {97}},\ \bibinfo {pages} {103014} (\bibinfo {year} {2018})},\ \Eprint
  {http://arxiv.org/abs/1711.07452} {arXiv:1711.07452 [astro-ph.HE]}
  \BibitemShut {NoStop}%
\bibitem [{\citenamefont {Samsing}\ and\ \citenamefont
  {D'Orazio}(2018)}]{Samsing:2018isx}%
  \BibitemOpen
  \bibfield  {author} {\bibinfo {author} {\bibfnamefont {J.}~\bibnamefont
  {Samsing}}\ and\ \bibinfo {author} {\bibfnamefont {D.~J.}\ \bibnamefont
  {D'Orazio}},\ }\href {\doibase 10.1093/mnras/sty2334} {\bibfield  {journal}
  {\bibinfo  {journal} {Mon. Not. Roy. Astron. Soc.}\ }\textbf {\bibinfo
  {volume} {481}},\ \bibinfo {pages} {5445} (\bibinfo {year} {2018})},\ \Eprint
  {http://arxiv.org/abs/1804.06519} {arXiv:1804.06519 [astro-ph.HE]}
  \BibitemShut {NoStop}%
\bibitem [{\citenamefont {Samsing}\ \emph {et~al.}(2020)\citenamefont
  {Samsing}, \citenamefont {D'Orazio}, \citenamefont {Kremer}, \citenamefont
  {Rodriguez},\ and\ \citenamefont {Askar}}]{Samsing:2019dtb}%
  \BibitemOpen
  \bibfield  {author} {\bibinfo {author} {\bibfnamefont {J.}~\bibnamefont
  {Samsing}}, \bibinfo {author} {\bibfnamefont {D.~J.}\ \bibnamefont
  {D'Orazio}}, \bibinfo {author} {\bibfnamefont {K.}~\bibnamefont {Kremer}},
  \bibinfo {author} {\bibfnamefont {C.~L.}\ \bibnamefont {Rodriguez}}, \ and\
  \bibinfo {author} {\bibfnamefont {A.}~\bibnamefont {Askar}},\ }\href
  {\doibase 10.1103/PhysRevD.101.123010} {\bibfield  {journal} {\bibinfo
  {journal} {Phys. Rev. D}\ }\textbf {\bibinfo {volume} {101}},\ \bibinfo
  {pages} {123010} (\bibinfo {year} {2020})},\ \Eprint
  {http://arxiv.org/abs/1907.11231} {arXiv:1907.11231 [astro-ph.HE]}
  \BibitemShut {NoStop}%
\bibitem [{\citenamefont {Antonini}\ and\ \citenamefont
  {Perets}(2012)}]{Antonini:2012ad}%
  \BibitemOpen
  \bibfield  {author} {\bibinfo {author} {\bibfnamefont {F.}~\bibnamefont
  {Antonini}}\ and\ \bibinfo {author} {\bibfnamefont {H.~B.}\ \bibnamefont
  {Perets}},\ }\href {\doibase 10.1088/0004-637X/757/1/27} {\bibfield
  {journal} {\bibinfo  {journal} {Astrophys. J.}\ }\textbf {\bibinfo {volume}
  {757}},\ \bibinfo {pages} {27} (\bibinfo {year} {2012})},\ \Eprint
  {http://arxiv.org/abs/1203.2938} {arXiv:1203.2938 [astro-ph.GA]} \BibitemShut
  {NoStop}%
\bibitem [{\citenamefont {Hoang}\ \emph {et~al.}(2018)\citenamefont {Hoang},
  \citenamefont {Naoz}, \citenamefont {Kocsis}, \citenamefont {Rasio},\ and\
  \citenamefont {Dosopoulou}}]{Hoang:2017fvh}%
  \BibitemOpen
  \bibfield  {author} {\bibinfo {author} {\bibfnamefont {B.-M.}\ \bibnamefont
  {Hoang}}, \bibinfo {author} {\bibfnamefont {S.}~\bibnamefont {Naoz}},
  \bibinfo {author} {\bibfnamefont {B.}~\bibnamefont {Kocsis}}, \bibinfo
  {author} {\bibfnamefont {F.~A.}\ \bibnamefont {Rasio}}, \ and\ \bibinfo
  {author} {\bibfnamefont {F.}~\bibnamefont {Dosopoulou}},\ }\href {\doibase
  10.3847/1538-4357/aaafce} {\bibfield  {journal} {\bibinfo  {journal}
  {Astrophys. J.}\ }\textbf {\bibinfo {volume} {856}},\ \bibinfo {pages} {140}
  (\bibinfo {year} {2018})},\ \Eprint {http://arxiv.org/abs/1706.09896}
  {arXiv:1706.09896 [astro-ph.HE]} \BibitemShut {NoStop}%
\bibitem [{\citenamefont {Fragione}\ and\ \citenamefont
  {Bromberg}(2019)}]{Fragione:2019dtr}%
  \BibitemOpen
  \bibfield  {author} {\bibinfo {author} {\bibfnamefont {G.}~\bibnamefont
  {Fragione}}\ and\ \bibinfo {author} {\bibfnamefont {O.}~\bibnamefont
  {Bromberg}},\ }\href {\doibase 10.1093/mnras/stz2024} {\bibfield  {journal}
  {\bibinfo  {journal} {Mon. Not. Roy. Astron. Soc.}\ }\textbf {\bibinfo
  {volume} {488}},\ \bibinfo {pages} {4370} (\bibinfo {year} {2019})},\ \Eprint
  {http://arxiv.org/abs/1903.09659} {arXiv:1903.09659 [astro-ph.GA]}
  \BibitemShut {NoStop}%
\bibitem [{\citenamefont {Randall}\ and\ \citenamefont
  {Xianyu}(2018{\natexlab{a}})}]{Randall:2017jop}%
  \BibitemOpen
  \bibfield  {author} {\bibinfo {author} {\bibfnamefont {L.}~\bibnamefont
  {Randall}}\ and\ \bibinfo {author} {\bibfnamefont {Z.-Z.}\ \bibnamefont
  {Xianyu}},\ }\href {\doibase 10.3847/1538-4357/aaa1a2} {\bibfield  {journal}
  {\bibinfo  {journal} {Astrophys. J.}\ }\textbf {\bibinfo {volume} {853}},\
  \bibinfo {pages} {93} (\bibinfo {year} {2018}{\natexlab{a}})},\ \Eprint
  {http://arxiv.org/abs/1708.08569} {arXiv:1708.08569 [gr-qc]} \BibitemShut
  {NoStop}%
\bibitem [{\citenamefont {Randall}\ and\ \citenamefont
  {Xianyu}(2018{\natexlab{b}})}]{Randall:2018nud}%
  \BibitemOpen
  \bibfield  {author} {\bibinfo {author} {\bibfnamefont {L.}~\bibnamefont
  {Randall}}\ and\ \bibinfo {author} {\bibfnamefont {Z.-Z.}\ \bibnamefont
  {Xianyu}},\ }\href {\doibase 10.3847/1538-4357/aad7fe} {\bibfield  {journal}
  {\bibinfo  {journal} {Astrophys. J.}\ }\textbf {\bibinfo {volume} {864}},\
  \bibinfo {pages} {134} (\bibinfo {year} {2018}{\natexlab{b}})},\ \Eprint
  {http://arxiv.org/abs/1802.05718} {arXiv:1802.05718 [gr-qc]} \BibitemShut
  {NoStop}%
\bibitem [{\citenamefont {Randall}\ and\ \citenamefont
  {Xianyu}(2019{\natexlab{a}})}]{Randall:2018lnh}%
  \BibitemOpen
  \bibfield  {author} {\bibinfo {author} {\bibfnamefont {L.}~\bibnamefont
  {Randall}}\ and\ \bibinfo {author} {\bibfnamefont {Z.-Z.}\ \bibnamefont
  {Xianyu}},\ }\href {\doibase 10.3847/1538-4357/ab20c6} {\bibfield  {journal}
  {\bibinfo  {journal} {Astrophys. J.}\ }\textbf {\bibinfo {volume} {878}},\
  \bibinfo {pages} {75} (\bibinfo {year} {2019}{\natexlab{a}})},\ \Eprint
  {http://arxiv.org/abs/1805.05335} {arXiv:1805.05335 [gr-qc]} \BibitemShut
  {NoStop}%
\bibitem [{\citenamefont {Silsbee}\ and\ \citenamefont
  {Tremaine}(2017)}]{Silsbee:2016djf}%
  \BibitemOpen
  \bibfield  {author} {\bibinfo {author} {\bibfnamefont {K.}~\bibnamefont
  {Silsbee}}\ and\ \bibinfo {author} {\bibfnamefont {S.}~\bibnamefont
  {Tremaine}},\ }\href {\doibase 10.3847/1538-4357/aa5729} {\bibfield
  {journal} {\bibinfo  {journal} {Astrophys. J.}\ }\textbf {\bibinfo {volume}
  {836}},\ \bibinfo {pages} {39} (\bibinfo {year} {2017})},\ \Eprint
  {http://arxiv.org/abs/1608.07642} {arXiv:1608.07642 [astro-ph.HE]}
  \BibitemShut {NoStop}%
\bibitem [{\citenamefont {Randall}\ and\ \citenamefont
  {Xianyu}(2019{\natexlab{b}})}]{Randall:2019sab}%
  \BibitemOpen
  \bibfield  {author} {\bibinfo {author} {\bibfnamefont {L.}~\bibnamefont
  {Randall}}\ and\ \bibinfo {author} {\bibfnamefont {Z.-Z.}\ \bibnamefont
  {Xianyu}},\ }\href@noop {} {\  (\bibinfo {year} {2019}{\natexlab{b}})},\
  \Eprint {http://arxiv.org/abs/1902.08604} {arXiv:1902.08604 [astro-ph.HE]}
  \BibitemShut {NoStop}%
\bibitem [{\citenamefont {Deme}\ \emph {et~al.}(2020)\citenamefont {Deme},
  \citenamefont {Hoang}, \citenamefont {Naoz},\ and\ \citenamefont
  {Kocsis}}]{Deme:2020ewx}%
  \BibitemOpen
  \bibfield  {author} {\bibinfo {author} {\bibfnamefont {B.}~\bibnamefont
  {Deme}}, \bibinfo {author} {\bibfnamefont {B.-M.}\ \bibnamefont {Hoang}},
  \bibinfo {author} {\bibfnamefont {S.}~\bibnamefont {Naoz}}, \ and\ \bibinfo
  {author} {\bibfnamefont {B.}~\bibnamefont {Kocsis}},\ }\href {\doibase
  10.3847/1538-4357/abafa3} {\bibfield  {journal} {\bibinfo  {journal}
  {Astrophys. J.}\ }\textbf {\bibinfo {volume} {901}},\ \bibinfo {pages} {125}
  (\bibinfo {year} {2020})},\ \Eprint {http://arxiv.org/abs/2005.03677}
  {arXiv:2005.03677 [astro-ph.HE]} \BibitemShut {NoStop}%
\bibitem [{\citenamefont {Ruan}\ \emph {et~al.}(2020)\citenamefont {Ruan},
  \citenamefont {Guo}, \citenamefont {Cai},\ and\ \citenamefont
  {Zhang}}]{Ruan:2018tsw}%
  \BibitemOpen
  \bibfield  {author} {\bibinfo {author} {\bibfnamefont {W.-H.}\ \bibnamefont
  {Ruan}}, \bibinfo {author} {\bibfnamefont {Z.-K.}\ \bibnamefont {Guo}},
  \bibinfo {author} {\bibfnamefont {R.-G.}\ \bibnamefont {Cai}}, \ and\
  \bibinfo {author} {\bibfnamefont {Y.-Z.}\ \bibnamefont {Zhang}},\ }\href
  {\doibase 10.1142/S0217751X2050075X} {\bibfield  {journal} {\bibinfo
  {journal} {Int. J. Mod. Phys. A}\ }\textbf {\bibinfo {volume} {35}},\
  \bibinfo {pages} {2050075} (\bibinfo {year} {2020})},\ \Eprint
  {http://arxiv.org/abs/1807.09495} {arXiv:1807.09495 [gr-qc]} \BibitemShut
  {NoStop}%
\bibitem [{\citenamefont {Liu}\ \emph {et~al.}(2020)\citenamefont {Liu},
  \citenamefont {Hu}, \citenamefont {Zhang},\ and\ \citenamefont
  {Mei}}]{Liu:2020eko}%
  \BibitemOpen
  \bibfield  {author} {\bibinfo {author} {\bibfnamefont {S.}~\bibnamefont
  {Liu}}, \bibinfo {author} {\bibfnamefont {Y.-M.}\ \bibnamefont {Hu}},
  \bibinfo {author} {\bibfnamefont {J.-d.}\ \bibnamefont {Zhang}}, \ and\
  \bibinfo {author} {\bibfnamefont {J.}~\bibnamefont {Mei}},\ }\href {\doibase
  10.1103/PhysRevD.101.103027} {\bibfield  {journal} {\bibinfo  {journal}
  {Phys. Rev. D}\ }\textbf {\bibinfo {volume} {101}},\ \bibinfo {pages}
  {103027} (\bibinfo {year} {2020})},\ \Eprint
  {http://arxiv.org/abs/2004.14242} {arXiv:2004.14242 [astro-ph.HE]}
  \BibitemShut {NoStop}%
\bibitem [{\citenamefont {Toubiana}\ \emph {et~al.}(2020)\citenamefont
  {Toubiana}, \citenamefont {Marsat}, \citenamefont {Babak}, \citenamefont
  {Baker},\ and\ \citenamefont {Dal~Canton}}]{Toubiana:2020cqv}%
  \BibitemOpen
  \bibfield  {author} {\bibinfo {author} {\bibfnamefont {A.}~\bibnamefont
  {Toubiana}}, \bibinfo {author} {\bibfnamefont {S.}~\bibnamefont {Marsat}},
  \bibinfo {author} {\bibfnamefont {S.}~\bibnamefont {Babak}}, \bibinfo
  {author} {\bibfnamefont {J.}~\bibnamefont {Baker}}, \ and\ \bibinfo {author}
  {\bibfnamefont {T.}~\bibnamefont {Dal~Canton}},\ }\href {\doibase
  10.1103/PhysRevD.102.124037} {\bibfield  {journal} {\bibinfo  {journal}
  {Phys. Rev. D}\ }\textbf {\bibinfo {volume} {102}},\ \bibinfo {pages}
  {124037} (\bibinfo {year} {2020})},\ \Eprint
  {http://arxiv.org/abs/2007.08544} {arXiv:2007.08544 [gr-qc]} \BibitemShut
  {NoStop}%
\bibitem [{\citenamefont {Buscicchio}\ \emph {et~al.}(2021)\citenamefont
  {Buscicchio}, \citenamefont {Klein}, \citenamefont {Roebber}, \citenamefont
  {Moore}, \citenamefont {Gerosa}, \citenamefont {Finch},\ and\ \citenamefont
  {Vecchio}}]{Buscicchio:2021dph}%
  \BibitemOpen
  \bibfield  {author} {\bibinfo {author} {\bibfnamefont {R.}~\bibnamefont
  {Buscicchio}}, \bibinfo {author} {\bibfnamefont {A.}~\bibnamefont {Klein}},
  \bibinfo {author} {\bibfnamefont {E.}~\bibnamefont {Roebber}}, \bibinfo
  {author} {\bibfnamefont {C.~J.}\ \bibnamefont {Moore}}, \bibinfo {author}
  {\bibfnamefont {D.}~\bibnamefont {Gerosa}}, \bibinfo {author} {\bibfnamefont
  {E.}~\bibnamefont {Finch}}, \ and\ \bibinfo {author} {\bibfnamefont
  {A.}~\bibnamefont {Vecchio}},\ }\href@noop {} {\  (\bibinfo {year} {2021})},\
  \Eprint {http://arxiv.org/abs/2106.05259} {arXiv:2106.05259 [astro-ph.HE]}
  \BibitemShut {NoStop}%
\bibitem [{\citenamefont {Randall}\ \emph {et~al.}(2021)\citenamefont
  {Randall}, \citenamefont {Shelest},\ and\ \citenamefont
  {Xianyu}}]{Randall:2021xjy}%
  \BibitemOpen
  \bibfield  {author} {\bibinfo {author} {\bibfnamefont {L.}~\bibnamefont
  {Randall}}, \bibinfo {author} {\bibfnamefont {A.}~\bibnamefont {Shelest}}, \
  and\ \bibinfo {author} {\bibfnamefont {Z.-Z.}\ \bibnamefont {Xianyu}},\
  }\href@noop {} {\  (\bibinfo {year} {2021})},\ \Eprint
  {http://arxiv.org/abs/2103.16030} {arXiv:2103.16030 [astro-ph.HE]}
  \BibitemShut {NoStop}%
\bibitem [{\citenamefont {Faucher-Giguere}\ and\ \citenamefont
  {Loeb}(2011)}]{Faucher-Giguere:2010dol}%
  \BibitemOpen
  \bibfield  {author} {\bibinfo {author} {\bibfnamefont {C.~A.}\ \bibnamefont
  {Faucher-Giguere}}\ and\ \bibinfo {author} {\bibfnamefont {A.}~\bibnamefont
  {Loeb}},\ }\href {\doibase 10.1111/j.1365-2966.2011.19019.x} {\bibfield
  {journal} {\bibinfo  {journal} {Mon. Not. Roy. Astron. Soc.}\ }\textbf
  {\bibinfo {volume} {415}},\ \bibinfo {pages} {3951} (\bibinfo {year}
  {2011})},\ \Eprint {http://arxiv.org/abs/1012.0573} {arXiv:1012.0573
  [astro-ph.HE]} \BibitemShut {NoStop}%
\bibitem [{\citenamefont {Shao}\ and\ \citenamefont {Li}(2018)}]{Shao:2018qpt}%
  \BibitemOpen
  \bibfield  {author} {\bibinfo {author} {\bibfnamefont {Y.}~\bibnamefont
  {Shao}}\ and\ \bibinfo {author} {\bibfnamefont {X.-D.}\ \bibnamefont {Li}},\
  }\href {\doibase 10.1093/mnrasl/sly063} {\bibfield  {journal} {\bibinfo
  {journal} {Mon. Not. Roy. Astron. Soc.}\ }\textbf {\bibinfo {volume} {477}},\
  \bibinfo {pages} {L128} (\bibinfo {year} {2018})},\ \Eprint
  {http://arxiv.org/abs/1804.06014} {arXiv:1804.06014 [astro-ph.HE]}
  \BibitemShut {NoStop}%
\bibitem [{\citenamefont {Barker}\ and\ \citenamefont
  {O'Connell}(1975)}]{Barker:1975ae}%
  \BibitemOpen
  \bibfield  {author} {\bibinfo {author} {\bibfnamefont {B.~M.}\ \bibnamefont
  {Barker}}\ and\ \bibinfo {author} {\bibfnamefont {R.~F.}\ \bibnamefont
  {O'Connell}},\ }\href {\doibase 10.1103/PhysRevD.12.329} {\bibfield
  {journal} {\bibinfo  {journal} {Phys. Rev. D}\ }\textbf {\bibinfo {volume}
  {12}},\ \bibinfo {pages} {329} (\bibinfo {year} {1975})}\BibitemShut
  {NoStop}%
\bibitem [{\citenamefont {Poisson}(1998)}]{Poisson:1997ha}%
  \BibitemOpen
  \bibfield  {author} {\bibinfo {author} {\bibfnamefont {E.}~\bibnamefont
  {Poisson}},\ }\href {\doibase 10.1103/PhysRevD.57.5287} {\bibfield  {journal}
  {\bibinfo  {journal} {Phys. Rev. D}\ }\textbf {\bibinfo {volume} {57}},\
  \bibinfo {pages} {5287} (\bibinfo {year} {1998})},\ \Eprint
  {http://arxiv.org/abs/gr-qc/9709032} {arXiv:gr-qc/9709032} \BibitemShut
  {NoStop}%
\bibitem [{\citenamefont {Barack}\ and\ \citenamefont
  {Cutler}(2007)}]{Barack:2006pq}%
  \BibitemOpen
  \bibfield  {author} {\bibinfo {author} {\bibfnamefont {L.}~\bibnamefont
  {Barack}}\ and\ \bibinfo {author} {\bibfnamefont {C.}~\bibnamefont
  {Cutler}},\ }\href {\doibase 10.1103/PhysRevD.75.042003} {\bibfield
  {journal} {\bibinfo  {journal} {Phys. Rev. D}\ }\textbf {\bibinfo {volume}
  {75}},\ \bibinfo {pages} {042003} (\bibinfo {year} {2007})},\ \Eprint
  {http://arxiv.org/abs/gr-qc/0612029} {arXiv:gr-qc/0612029} \BibitemShut
  {NoStop}%
\bibitem [{\citenamefont {Krishnendu}\ \emph {et~al.}(2017)\citenamefont
  {Krishnendu}, \citenamefont {Arun},\ and\ \citenamefont
  {Mishra}}]{Krishnendu:2017shb}%
  \BibitemOpen
  \bibfield  {author} {\bibinfo {author} {\bibfnamefont {N.~V.}\ \bibnamefont
  {Krishnendu}}, \bibinfo {author} {\bibfnamefont {K.~G.}\ \bibnamefont
  {Arun}}, \ and\ \bibinfo {author} {\bibfnamefont {C.~K.}\ \bibnamefont
  {Mishra}},\ }\href {\doibase 10.1103/PhysRevLett.119.091101} {\bibfield
  {journal} {\bibinfo  {journal} {Phys. Rev. Lett.}\ }\textbf {\bibinfo
  {volume} {119}},\ \bibinfo {pages} {091101} (\bibinfo {year} {2017})},\
  \Eprint {http://arxiv.org/abs/1701.06318} {arXiv:1701.06318 [gr-qc]}
  \BibitemShut {NoStop}%
\bibitem [{\citenamefont {Ferreira}\ \emph {et~al.}(2017)\citenamefont
  {Ferreira}, \citenamefont {Macedo},\ and\ \citenamefont
  {Cardoso}}]{Ferreira:2017pth}%
  \BibitemOpen
  \bibfield  {author} {\bibinfo {author} {\bibfnamefont {M.~C.}\ \bibnamefont
  {Ferreira}}, \bibinfo {author} {\bibfnamefont {C.~F.~B.}\ \bibnamefont
  {Macedo}}, \ and\ \bibinfo {author} {\bibfnamefont {V.}~\bibnamefont
  {Cardoso}},\ }\href {\doibase 10.1103/PhysRevD.96.083017} {\bibfield
  {journal} {\bibinfo  {journal} {Phys. Rev. D}\ }\textbf {\bibinfo {volume}
  {96}},\ \bibinfo {pages} {083017} (\bibinfo {year} {2017})},\ \Eprint
  {http://arxiv.org/abs/1710.00830} {arXiv:1710.00830 [gr-qc]} \BibitemShut
  {NoStop}%
\bibitem [{\citenamefont {Hannuksela}\ \emph {et~al.}(2019)\citenamefont
  {Hannuksela}, \citenamefont {Wong}, \citenamefont {Brito}, \citenamefont
  {Berti},\ and\ \citenamefont {Li}}]{Hannuksela:2018izj}%
  \BibitemOpen
  \bibfield  {author} {\bibinfo {author} {\bibfnamefont {O.~A.}\ \bibnamefont
  {Hannuksela}}, \bibinfo {author} {\bibfnamefont {K.~W.~K.}\ \bibnamefont
  {Wong}}, \bibinfo {author} {\bibfnamefont {R.}~\bibnamefont {Brito}},
  \bibinfo {author} {\bibfnamefont {E.}~\bibnamefont {Berti}}, \ and\ \bibinfo
  {author} {\bibfnamefont {T.~G.~F.}\ \bibnamefont {Li}},\ }\href {\doibase
  10.1038/s41550-019-0712-4} {\bibfield  {journal} {\bibinfo  {journal} {Nature
  Astron.}\ }\textbf {\bibinfo {volume} {3}},\ \bibinfo {pages} {447} (\bibinfo
  {year} {2019})},\ \Eprint {http://arxiv.org/abs/1804.09659} {arXiv:1804.09659
  [astro-ph.HE]} \BibitemShut {NoStop}%
\bibitem [{\citenamefont {Zhang}\ and\ \citenamefont
  {Yang}(2020)}]{Zhang:2019eid}%
  \BibitemOpen
  \bibfield  {author} {\bibinfo {author} {\bibfnamefont {J.}~\bibnamefont
  {Zhang}}\ and\ \bibinfo {author} {\bibfnamefont {H.}~\bibnamefont {Yang}},\
  }\href {\doibase 10.1103/PhysRevD.101.043020} {\bibfield  {journal} {\bibinfo
   {journal} {Phys. Rev. D}\ }\textbf {\bibinfo {volume} {101}},\ \bibinfo
  {pages} {043020} (\bibinfo {year} {2020})},\ \Eprint
  {http://arxiv.org/abs/1907.13582} {arXiv:1907.13582 [gr-qc]} \BibitemShut
  {NoStop}%
\bibitem [{\citenamefont {Amorim}\ \emph {et~al.}(2019)\citenamefont {Amorim}
  \emph {et~al.}}]{GRAVITY:2019tuf}%
  \BibitemOpen
  \bibfield  {author} {\bibinfo {author} {\bibfnamefont {A.}~\bibnamefont
  {Amorim}} \emph {et~al.} (\bibinfo {collaboration} {GRAVITY}),\ }\href
  {\doibase 10.1093/mnras/stz2300} {\bibfield  {journal} {\bibinfo  {journal}
  {Mon. Not. Roy. Astron. Soc.}\ }\textbf {\bibinfo {volume} {489}},\ \bibinfo
  {pages} {4606} (\bibinfo {year} {2019})},\ \Eprint
  {http://arxiv.org/abs/1908.06681} {arXiv:1908.06681 [astro-ph.GA]}
  \BibitemShut {NoStop}%
\bibitem [{\citenamefont {Su}\ \emph {et~al.}()\citenamefont {Su},
  \citenamefont {Xianyu},\ and\ \citenamefont {Zhang}}]{preclong}%
  \BibitemOpen
  \bibfield  {author} {\bibinfo {author} {\bibfnamefont {B.}~\bibnamefont
  {Su}}, \bibinfo {author} {\bibfnamefont {Z.-Z.}\ \bibnamefont {Xianyu}}, \
  and\ \bibinfo {author} {\bibfnamefont {X.}~\bibnamefont {Zhang}},\
  }\href@noop {} {\enquote {\bibinfo {title} {to appear},}\ }\BibitemShut
  {NoStop}%
\bibitem [{\citenamefont {Yoshino}\ and\ \citenamefont
  {Kodama}(2014)}]{Yoshino:2013ofa}%
  \BibitemOpen
  \bibfield  {author} {\bibinfo {author} {\bibfnamefont {H.}~\bibnamefont
  {Yoshino}}\ and\ \bibinfo {author} {\bibfnamefont {H.}~\bibnamefont
  {Kodama}},\ }\href {\doibase 10.1093/ptep/ptu029} {\bibfield  {journal}
  {\bibinfo  {journal} {PTEP}\ }\textbf {\bibinfo {volume} {2014}},\ \bibinfo
  {pages} {043E02} (\bibinfo {year} {2014})},\ \Eprint
  {http://arxiv.org/abs/1312.2326} {arXiv:1312.2326 [gr-qc]} \BibitemShut
  {NoStop}%
\bibitem [{\citenamefont {Cardoso}\ \emph {et~al.}(2020)\citenamefont
  {Cardoso}, \citenamefont {Duque},\ and\ \citenamefont
  {Ikeda}}]{Cardoso:2020hca}%
  \BibitemOpen
  \bibfield  {author} {\bibinfo {author} {\bibfnamefont {V.}~\bibnamefont
  {Cardoso}}, \bibinfo {author} {\bibfnamefont {F.}~\bibnamefont {Duque}}, \
  and\ \bibinfo {author} {\bibfnamefont {T.}~\bibnamefont {Ikeda}},\ }\href
  {\doibase 10.1103/PhysRevD.101.064054} {\bibfield  {journal} {\bibinfo
  {journal} {Phys. Rev. D}\ }\textbf {\bibinfo {volume} {101}},\ \bibinfo
  {pages} {064054} (\bibinfo {year} {2020})},\ \Eprint
  {http://arxiv.org/abs/2001.01729} {arXiv:2001.01729 [gr-qc]} \BibitemShut
  {NoStop}%
\bibitem [{\citenamefont {Takahashi}\ and\ \citenamefont
  {Tanaka}(2021)}]{Takahashi:2021eso}%
  \BibitemOpen
  \bibfield  {author} {\bibinfo {author} {\bibfnamefont {T.}~\bibnamefont
  {Takahashi}}\ and\ \bibinfo {author} {\bibfnamefont {T.}~\bibnamefont
  {Tanaka}},\ }\href@noop {} {\  (\bibinfo {year} {2021})},\ \Eprint
  {http://arxiv.org/abs/2106.08836} {arXiv:2106.08836 [gr-qc]} \BibitemShut
  {NoStop}%
\bibitem [{\citenamefont {De~Luca}\ and\ \citenamefont
  {Pani}(2021)}]{DeLuca:2021ite}%
  \BibitemOpen
  \bibfield  {author} {\bibinfo {author} {\bibfnamefont {V.}~\bibnamefont
  {De~Luca}}\ and\ \bibinfo {author} {\bibfnamefont {P.}~\bibnamefont {Pani}},\
  }\href@noop {} {\  (\bibinfo {year} {2021})},\ \Eprint
  {http://arxiv.org/abs/2106.14428} {arXiv:2106.14428 [gr-qc]} \BibitemShut
  {NoStop}%
\bibitem [{\citenamefont {Einstein}\ \emph {et~al.}(1938)\citenamefont
  {Einstein}, \citenamefont {Infeld},\ and\ \citenamefont
  {Hoffmann}}]{Einstein:1938yz}%
  \BibitemOpen
  \bibfield  {author} {\bibinfo {author} {\bibfnamefont {A.}~\bibnamefont
  {Einstein}}, \bibinfo {author} {\bibfnamefont {L.}~\bibnamefont {Infeld}}, \
  and\ \bibinfo {author} {\bibfnamefont {B.}~\bibnamefont {Hoffmann}},\ }\href
  {\doibase 10.2307/1968714} {\bibfield  {journal} {\bibinfo  {journal} {Annals
  Math.}\ }\textbf {\bibinfo {volume} {39}},\ \bibinfo {pages} {65} (\bibinfo
  {year} {1938})}\BibitemShut {NoStop}%
\bibitem [{\citenamefont {Barack}\ and\ \citenamefont
  {Cutler}(2004)}]{Barack:2003fp}%
  \BibitemOpen
  \bibfield  {author} {\bibinfo {author} {\bibfnamefont {L.}~\bibnamefont
  {Barack}}\ and\ \bibinfo {author} {\bibfnamefont {C.}~\bibnamefont
  {Cutler}},\ }\href {\doibase 10.1103/PhysRevD.69.082005} {\bibfield
  {journal} {\bibinfo  {journal} {Phys. Rev.}\ }\textbf {\bibinfo {volume}
  {D69}},\ \bibinfo {pages} {082005} (\bibinfo {year} {2004})},\ \Eprint
  {http://arxiv.org/abs/gr-qc/0310125} {arXiv:gr-qc/0310125 [gr-qc]}
  \BibitemShut {NoStop}%
\bibitem [{\citenamefont {Klein}\ \emph {et~al.}(2016)\citenamefont {Klein}
  \emph {et~al.}}]{Klein:2015hvg}%
  \BibitemOpen
  \bibfield  {author} {\bibinfo {author} {\bibfnamefont {A.}~\bibnamefont
  {Klein}} \emph {et~al.},\ }\href {\doibase 10.1103/PhysRevD.93.024003}
  {\bibfield  {journal} {\bibinfo  {journal} {Phys. Rev.}\ }\textbf {\bibinfo
  {volume} {D93}},\ \bibinfo {pages} {024003} (\bibinfo {year} {2016})},\
  \Eprint {http://arxiv.org/abs/1511.05581} {arXiv:1511.05581 [gr-qc]}
  \BibitemShut {NoStop}%
\bibitem [{\citenamefont {Peters}(1964)}]{Peters:1964zz}%
  \BibitemOpen
  \bibfield  {author} {\bibinfo {author} {\bibfnamefont {P.~C.}\ \bibnamefont
  {Peters}},\ }\href {\doibase 10.1103/PhysRev.136.B1224} {\bibfield  {journal}
  {\bibinfo  {journal} {Phys. Rev.}\ }\textbf {\bibinfo {volume} {136}},\
  \bibinfo {pages} {B1224} (\bibinfo {year} {1964})}\BibitemShut {NoStop}%
\bibitem [{\citenamefont {Stairs}(2003)}]{Stairs:2003eg}%
  \BibitemOpen
  \bibfield  {author} {\bibinfo {author} {\bibfnamefont {I.~H.}\ \bibnamefont
  {Stairs}},\ }\href {\doibase 10.12942/lrr-2003-5} {\bibfield  {journal}
  {\bibinfo  {journal} {Living Rev. Rel.}\ }\textbf {\bibinfo {volume} {6}},\
  \bibinfo {pages} {5} (\bibinfo {year} {2003})},\ \Eprint
  {http://arxiv.org/abs/astro-ph/0307536} {arXiv:astro-ph/0307536} \BibitemShut
  {NoStop}%
\bibitem [{\citenamefont {Ikeda}\ \emph {et~al.}(2021)\citenamefont {Ikeda},
  \citenamefont {Bernard}, \citenamefont {Cardoso},\ and\ \citenamefont
  {Zilh\~ao}}]{Ikeda:2020xvt}%
  \BibitemOpen
  \bibfield  {author} {\bibinfo {author} {\bibfnamefont {T.}~\bibnamefont
  {Ikeda}}, \bibinfo {author} {\bibfnamefont {L.}~\bibnamefont {Bernard}},
  \bibinfo {author} {\bibfnamefont {V.}~\bibnamefont {Cardoso}}, \ and\
  \bibinfo {author} {\bibfnamefont {M.}~\bibnamefont {Zilh\~ao}},\ }\href
  {\doibase 10.1103/PhysRevD.103.024020} {\bibfield  {journal} {\bibinfo
  {journal} {Phys. Rev. D}\ }\textbf {\bibinfo {volume} {103}},\ \bibinfo
  {pages} {024020} (\bibinfo {year} {2021})},\ \Eprint
  {http://arxiv.org/abs/2010.00008} {arXiv:2010.00008 [gr-qc]} \BibitemShut
  {NoStop}%
\bibitem [{\citenamefont {Liu}\ and\ \citenamefont {Lyu}(2021)}]{Liu:2021llm}%
  \BibitemOpen
  \bibfield  {author} {\bibinfo {author} {\bibfnamefont {T.}~\bibnamefont
  {Liu}}\ and\ \bibinfo {author} {\bibfnamefont {K.-F.}\ \bibnamefont {Lyu}},\
  }\href@noop {} {\  (\bibinfo {year} {2021})},\ \Eprint
  {http://arxiv.org/abs/2107.09971} {arXiv:2107.09971 [astro-ph.HE]}
  \BibitemShut {NoStop}%
\end{thebibliography}
\end{document}